\documentclass[fleqn,usenatbib]{mnras}

\usepackage{newtxtext,newtxmath}

\usepackage[T1]{fontenc}
\usepackage{ae,aecompl}
\usepackage{soul}

\usepackage{graphicx}	
\usepackage{amsmath}	
\usepackage[usenames]{xcolor}
\usepackage{natbib}
\usepackage{hyperref}
\usepackage{mathtools}
\usepackage{xfrac}

\def\equationautorefname~#1\null{equation~(#1)}

\DeclareMathAlphabet{\mathpzc}{OT1}{pzc}{m}{it}\definecolor{purple}{RGB}{160,32,240}

\newcommand{\rev}[1]{\textcolor{black}{#1}}

\newcommand{\Mstar}{M_{\star}}
\newcommand{\Mearth}{M_{\oplus}}
\newcommand{\Mj}{M_{\rm J}}

\newcommand{\ta}{t_a}
\newcommand{\te}{t_e}

\newcommand{\efone}{e_{\rm free,1}}
\newcommand{\eftwo}{e_{\rm free,2}}
\newcommand{\wfone}{\varpi_{\rm free,1}}
\newcommand{\wftwo}{\varpi_{\rm free,2}}
\newcommand{\Phit}{\Phi_{\rm TTV}}

\newcommand{\dd}{\mathcal{D}/\Delta}
\newcommand{\mdd}{\mathrm{max}(\mathcal{D}/\Delta)}
\newcommand{\CA}{C_{\rm A}}
\newcommand{\CB}{C_{\rm B}}

\title[TTV phases]{Exciting the TTV Phases of Resonant Sub-Neptunes}

\author[Choksi \& Chiang]{Nick Choksi$^{1 \href{https://orcid.org/0000-0003-0690-1056}{\includegraphics[scale=0.4]{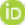}}}$\thanks{E-mail: nchoksi@berkeley.edu} and
Eugene Chiang$^{1,2\href{https://orcid.org/0000-0002-6246-2310}{\includegraphics[scale=0.4]{figures/orcid.pdf}}}$
\\
$^{1}$Astronomy Department, Theoretical Astrophysics Center, and Center for Integrative Planetary Science, University of California\\
\hspace{0.015in} Berkeley, Berkeley, CA 94720, USA\\
$^{2}$Department of Earth and Planetary Science, University of California, Berkeley, CA 94720, USA
}

\date{Released \today}

\pubyear{2022}

\begin{document}
\label{firstpage}
\pagerange{\pageref{firstpage}--\pageref{lastpage}}
\maketitle

\begin{abstract}
There are excesses of
sub-Neptunes just wide of period commensurabilities like the 3:2 and 2:1, and corresponding deficits narrow of them. Any theory that explains this period ratio structure must also explain the strong transit timing variations (TTVs) observed near resonance. Besides an amplitude and a period, a sinusoidal TTV has a phase. Often overlooked, TTV phases are effectively integration constants, encoding information about initial conditions or the environment. Many TTVs near resonance exhibit non-zero phases. This observation is surprising because dissipative processes that capture planets into resonance also damp TTV phases to zero. We show how both the period ratio structure and the non-zero TTV phases can be reproduced if pairs of sub-Neptunes capture into resonance in a gas disc while accompanied by a third \rev{eccentric} non-resonant body. Convergent migration and eccentricity damping \rev{by the disc} drives pairs to orbital period ratios wide of commensurability; then, after the disc clears, secular forcing by the third body phase-shifts the TTVs. The scenario predicts that resonant planets are apsidally aligned and possess eccentricities up to an order of magnitude larger than previously thought.
\end{abstract}



\section{Introduction}
\label{sec:Intro}

\begin{figure*}
\includegraphics[width=\textwidth]{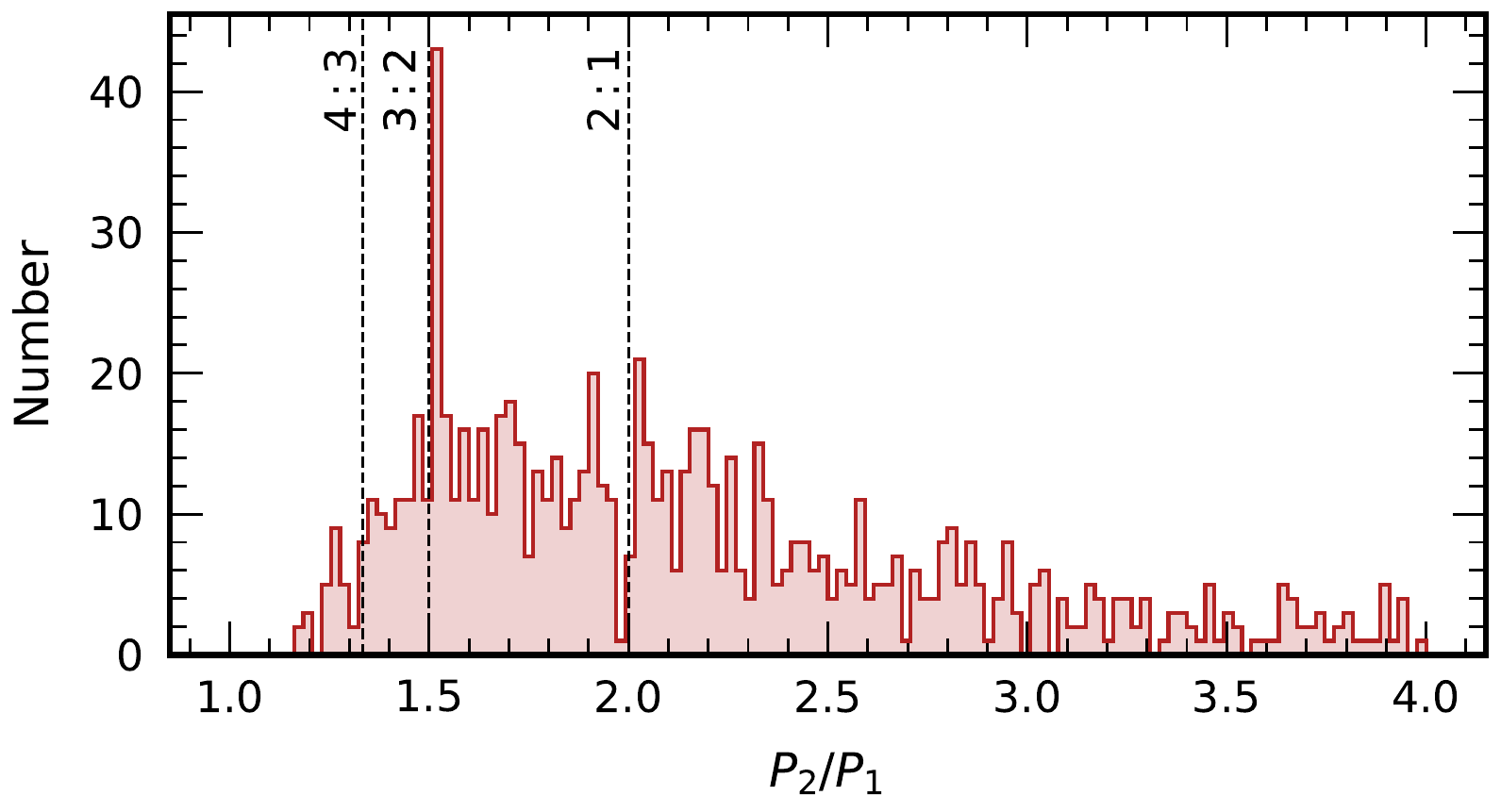}
\caption{Histogram of period ratios $P_2/P_1$ of neighboring sub-Neptunes (radii $< 4 R_{\oplus}$) in the NASA Exoplanet Archive as of May 2022. There is an excess of systems at period ratios a percent or so wide of the 3:2 commensurability, and a deficit of systems just short of it. The 2:1 and 4:3 period ratios display similar structure. Our goal is to find a theory that can explain not only these resonant ``peaks'' and ``troughs'', but also the phases of observed transit timing variations (TTVs), as shown in Figures \ref{fig:ttv_data2} and \ref{fig:data_3pan}.
}
  \label{fig:period_ratio_distributions}
\end{figure*}

Sub-Neptunes, planets with radii $\lesssim 4R_{\oplus}$, abound within an au of solar-type and low-mass stars \citep[e.g.,][]{fressin_etal_2013, dressing_charbonneau_2015, petigura_etal_2018, zhu_etal_2018, sandford_etal_2019, otegi_etal_2021, daylan_etal_2021, turtelboom_etal_2022}. They are frequently members of multi-planet systems.

Figure \ref{fig:period_ratio_distributions} shows the period ratios $P_2/P_1$ of  neighboring 
sub-Neptunes (subscript ``1'' for the inner planet and ``2'' for the outer), compiled from the NASA Exoplanet Archive. There appear to be excess numbers of pairs with commensurable periods,
most obviously near $P_2/P_1 =$ 3:2, and arguably also near 2:1 and 4:3.
These ``peaks'' in the histogram actually occur at $P_2/P_1$ values slightly greater than small-integer ratios. In terms of
\begin{equation}
    \Delta = \frac{q}{q+1}\frac{P_2}{P_1} - 1 \,,
\end{equation}
which measures a pair's fractional distance from a $(q+1)$:$q$ commensurability ($q$ is a positive integer), the observed 3:2, 2:1, and 4:3 peaks lie at $\Delta \sim +1\%$. Just adjacent to the peaks are population deficits or ``troughs'' at $\Delta \sim -1\%$, with the clearest trough situated shortward of 2:1 \citep{lissauer_etal_2011, fabrycky_etal_2014, steffen_hwang_2015}. These peak-trough features can be discerned out to orbital periods of tens of days \citep{choksi_chiang_2020}.

The peaks and troughs are signatures of capture into mean-motion resonance (for an introduction, see \citealt{murray_dermott_1999}). Capture is effected by slow changes in a planet pair's semimajor axes that bring period ratios nearer to commensurability. The migration is driven by some source of dissipation, e.g. the dissipation of tides raised on planets by their host stars \citep{lithwick_wu_2012, batygin_morbidelli_2013, millholland_laughlin_2019}, or the dissipation of waves excited by planets in their natal gas discs, a.k.a. dynamical friction \citep[e.g.][]{goldreich_tremaine_1980, lee_peale_2002}. The 3:2 and 2:1 peak-trough statistics can be reproduced in a scenario, staged late in a protoplanetary disc's life, whereby some sub-Neptune pairs migrate convergently into the peak (so that $P_2/P_1$ approaches commensurability from above), while others migrate divergently, out of the trough, because of eccentricity damping by dynamical friction \citep{choksi_chiang_2020, macdonald_etal_2020}. The observed preference of first-order resonances for $\Delta > 0$ is a consequence of dissipation driving systems to their fixed points, which are located intrinsically at $\Delta > 0$ (\citealt{goldreich_1965}; see also the introduction of \citealt{choksi_chiang_2020}). 

\begin{figure*}
\centering 
\includegraphics[width=\textwidth]{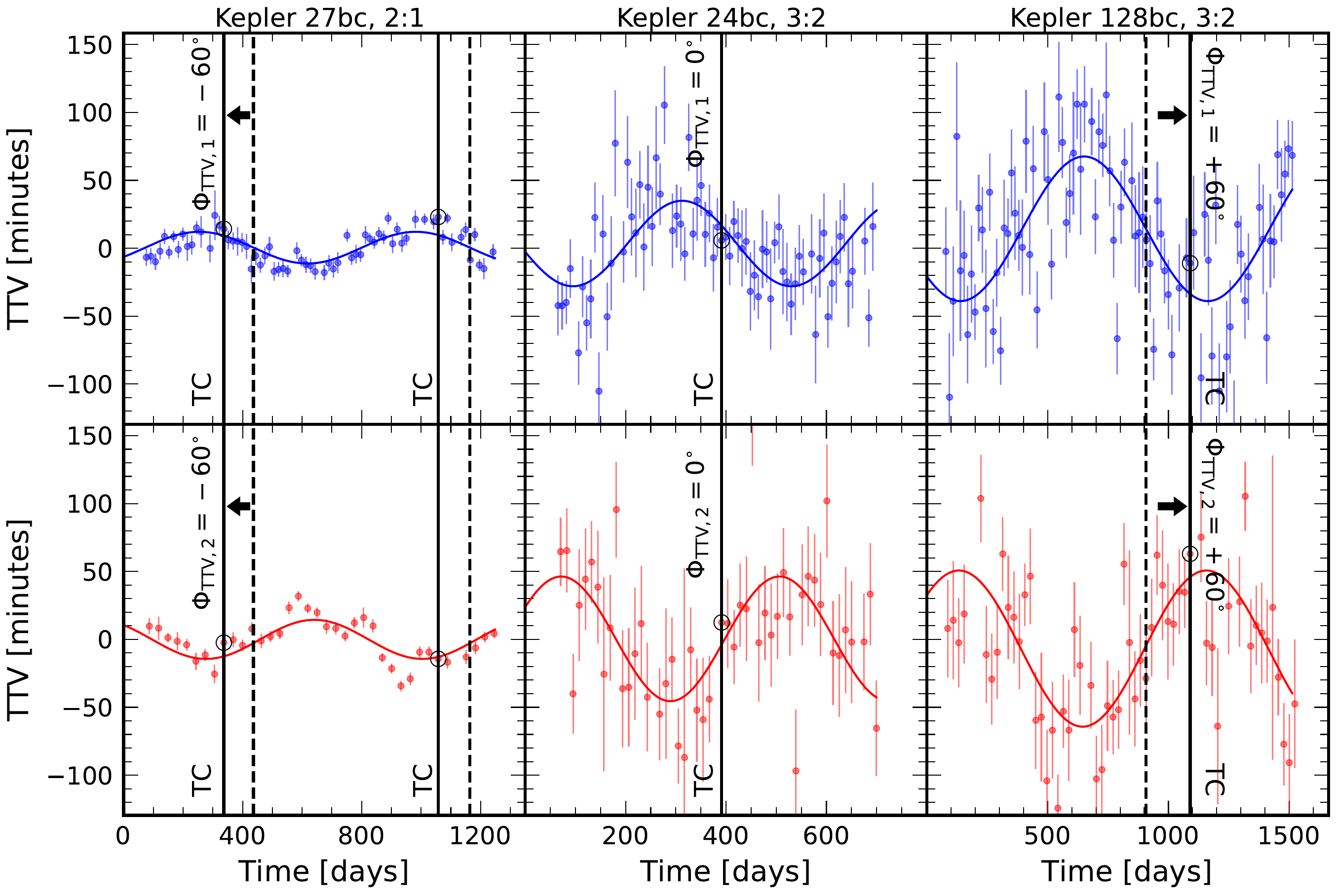}
  \caption{Example transit timing variations (TTVs) taken from \citet{rowe_etal_2015} for planets near the 2:1 and 3:2 resonances, fitted with sine waves. Each column showcases one resonant pair, with TTVs for the inner member plotted on top, and for the outer member on the bottom. Solid vertical lines and open circles mark ``transiting conjunctions'' (TCs) when both planets in the pair transit the star at nearly the same time. 
  These are 
  identified by calculating when the two planets have mid-transit times that differ by less than the time it takes the inner planet to cross the host star, $2R_{\odot}/v_{\rm K,1} \approx 5\,\mathrm{hr}\,(P_1/20\,\rm d)^{1/3}$, evaluated for a circular orbit of velocity $v_{\rm K,1}$ and period $P_1$ around a solar-mass star. Transiting conjunctions occur once per TTV cycle and thus offer a reference time against which to measure the TTV phase $\Phit$. We define $\Phit$ as the phase difference between the TC and the nearest TTV zero-crossing (dashed line), either the nearest descending crossing for the inner planet, or the nearest ascending crossing for the outer planet. If the TC occurs at the same time as the TTV sine wave crosses zero, then $\Phit = 0$ (middle column). If the TC precedes the nearest zero-crossing, then $\Phit < 0$ (left column); otherwise $\Phi_{\rm TTV} > 0$ (right column).
  }
    \label{fig:ttv_data2}
\end{figure*}

\begin{figure*}
\includegraphics[width=\textwidth]{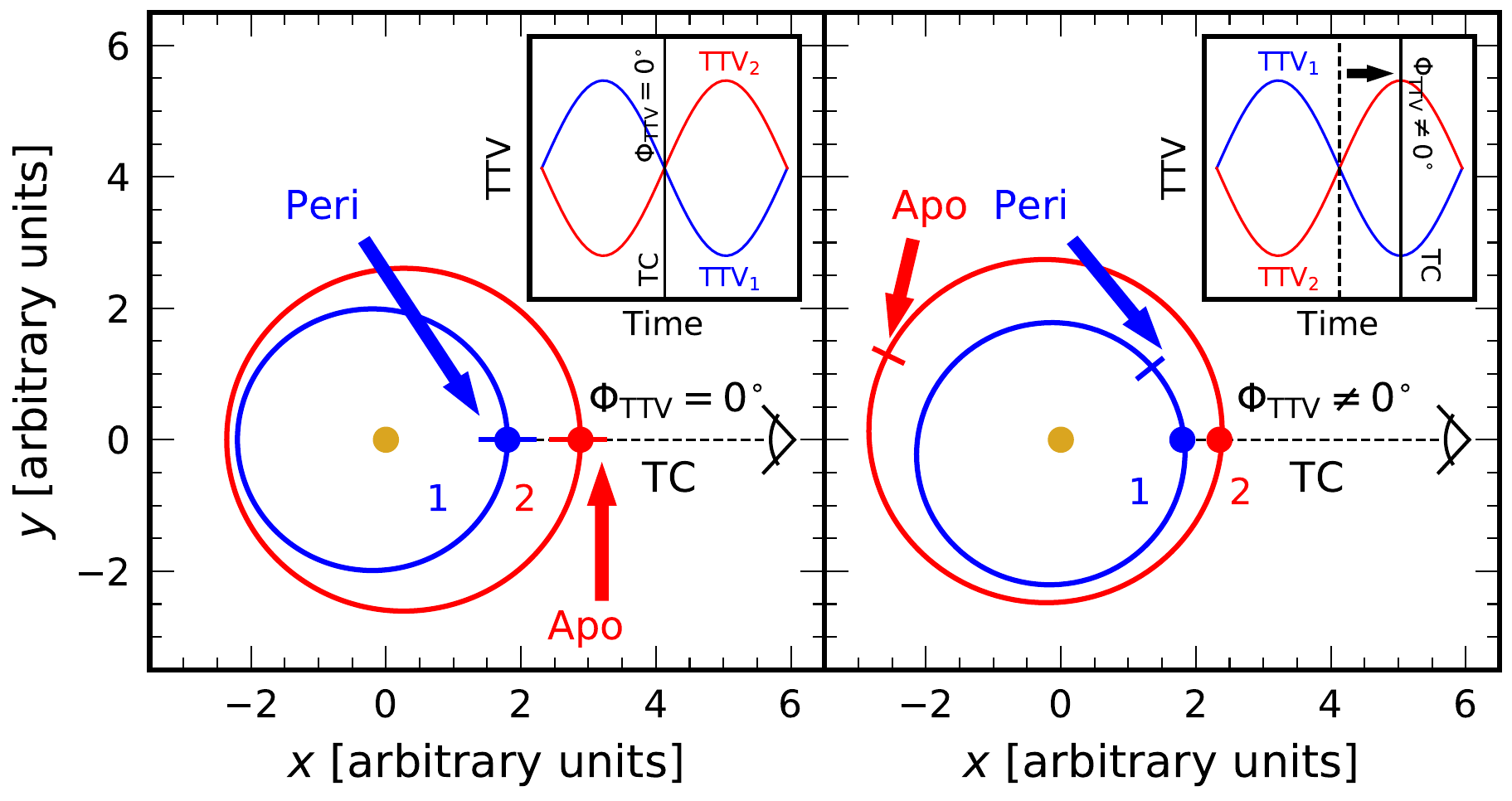}
\caption{Geometric interpretation of TTV phases. \textit{Left:} When $\Phit = 0$ for both planets, the inner planet transits at its periapse and the outer planet at its apoapse during a transiting conjunction. This orbital configuration corresponds to the fixed point of the resonance. Standard migration scenarios for an isolated pair of planets capturing into resonance drive the system to this fixed point and therefore predict $\Phit=0$.
\textit{Right:} A non-zero $\Phit$ implies the planets are not at peri/apo during a transiting conjunction; the system has significant free eccentricities and has wandered away from the fixed point of the resonance. Orbits and corresponding TTVs are drawn from simulated resonant systems (they are not schematic).
}
  \label{fig:cartoon}
\end{figure*}

\begin{figure*}
\centering 
\includegraphics[width=1.05\textwidth]{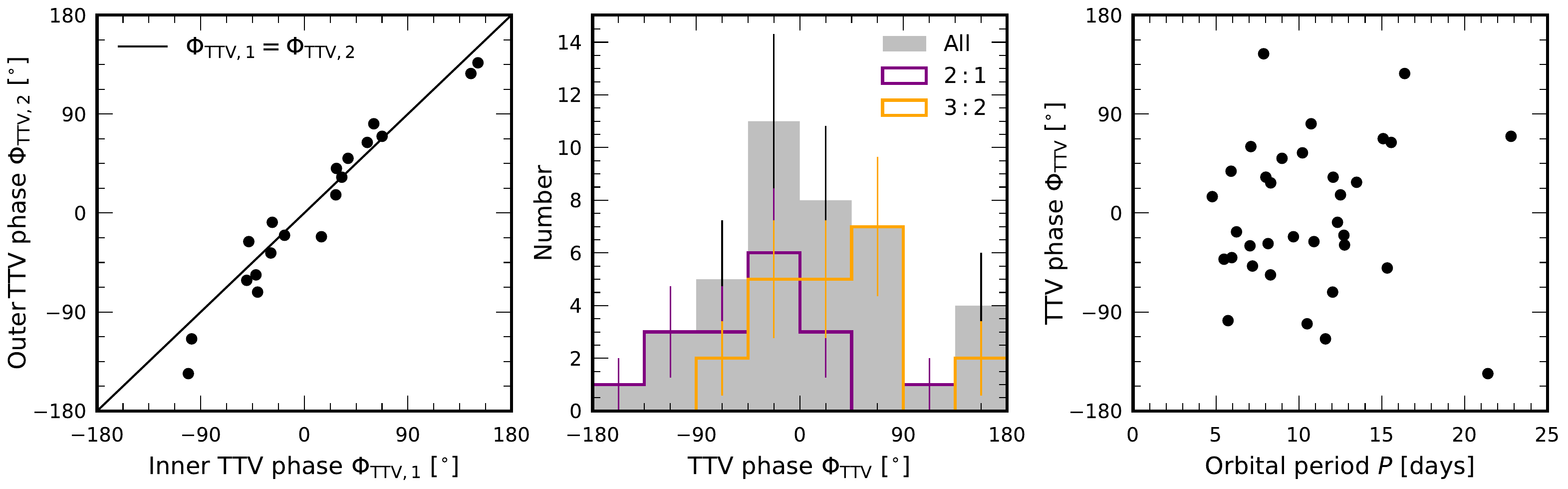}
\caption{ \rev{Observed TTV phases of planet pairs in the 3:2 and 2:1 peaks of Fig.~\ref{fig:period_ratio_distributions} (timing data from \citealt{rowe_etal_2015}). The data shown here are for TTVs which
appear sinusoidal and whose super-periods match to within 20\% that of a near-resonant two-planet system, $P_{\rm super} = P_1/\left(q |\Delta|\right)$.
}
\textit{Left:} For most such pairs, 
the TTVs of the inner member are anti-correlated with those of the outer member (Fig.~\ref{fig:ttv_data2}), and so 
$\Phi_{\rm TTV,1} \approx \Phi_{\rm TTV,2}$. 
\textit{Middle:} Histograms of TTV phases, now without distinguishing between the inner and outer phases. Standard migration scenarios for isolated pairs of planets capturing into resonance predict $\Phi_{\rm TTV} = 0$. Although the observed distribution of phases peaks near zero, many systems exhibit non-zero phases. \textit{Right:} TTV phases evince no dependence on planetary orbital periods $P$.
}
\label{fig:data_3pan}
\end{figure*}

The orbital periods in Figure \ref{fig:period_ratio_distributions} are evaluated using the average time between a given planet's transits, as measured from transit surveys like the {\it Kepler} and {\it TESS} space missions. The actual time between transits can vary from transit to transit --- the transit timing variation (TTV) is the fluctuation about the average. 
The closer a pair of planets is to a period commensurability (i.e. the smaller is $\Delta$), the larger are its TTVs \citep[e.g.][]{agol_etal_2005, holman_murray_2005}.\footnote{We distinguish in this paper between ``commensurability'' and ``resonance''. The quantity $\Delta$ measures the distance to a $(q+1)$:$q$ period commensurability, and is distinct from the resonant arguments (e.g. \citealt{murray_dermott_1999}) used to decide whether a system is librating in resonance, or circulating out of it. When we say a system is ``in'' or ``captured into'' or ``locked in'' resonance, we mean that at least one resonant argument is librating. Our usage differs from some of the TTV literature, which instead uses $\Delta$ to decide whether a system is near or in a resonance (e.g. \citealt{agol_etal_2005}), and treats commensurability and resonance as synonymous. \label{foot:res}}
Many of the sub-Neptune pairs in the 3:2 and 2:1 resonant peaks exhibit simple sinusoidal TTV cycles; some examples from the {\it Kepler} TTV catalogue of \citet{rowe_etal_2015}  are illustrated in Figure \ref{fig:ttv_data2} (see also Table 2 of \citealt{hadden_lithwick_2017}). Sinusoidal TTVs are expected for two bodies near commensurability 
\citep{lithwick_etal_2012, hadden_lithwick_2016,  deck_agol_2016, nesvorny_2016}, 
with the TTVs of the inner and outer members varying 
over the same TTV ``super-period'', $P_{\rm super} \sim P_1/|\Delta|$. For the examples shown in Fig.~\ref{fig:ttv_data2},
the inner and outer planet TTVs are anti-correlated: when one rises, the other falls. The TTV amplitudes depend on planet masses and eccentricities \citep{lithwick_etal_2012, wu_lithwick_2013, hadden_lithwick_2014, hadden_lithwick_2016, hadden_lithwick_2017}, and for the systems of interest here are on the order of an hour \citep{mazeh_etal_2013, rowe_etal_2015}.

A TTV cycle is described not only by its amplitude and period but also by its phase 
\citep{lithwick_etal_2012, deck_agol_2016}. A simple working definition for the TTV phase starts by identifying a transiting conjunction (TC). A TC is a conjunction (occurring when the star and the two planets lie along a line) that coincides with a transit. Equivalently, TCs occur when both planets transit simultaneously. Such times are marked ``TC'' in Fig.~\ref{fig:ttv_data2}; they occur once per super-period. 
The phase $\Phi_{\rm TTV}$ is the phase interval between a TC and the nearest TTV zero-crossing. We use the ``descending'' zero-crossing (when the TTV crosses zero from above) when evaluating the phase $\Phi_{\rm TTV,1}$ of the inner member of a resonant pair, and the ``ascending'' zero-crossing for the outer phase $\Phi_{\rm TTV,2}$. 
For anti-correlated TTV pairs like those in Fig.~\ref{fig:ttv_data2}, this convention implies $\Phi_{\rm TTV,1} = \Phi_{\rm TTV,2}$,\footnote{Our convention differs from that of \cite{lithwick_etal_2012} who measure the phases of both the inner and outer planet's TTV relative to the descending zero-crossing.} in which case a single phase $\Phi_{\rm TTV}$ uniquely describes the system. The resonant pair featured in the middle column of Fig.~\ref{fig:ttv_data2} has $\Phi_{\rm TTV} = 0$ for both planets because the TCs coincide with the TTV zero-crossings. If the TC precedes the nearest zero-crossing, then $\Phi_{\rm TTV} < 0$ (left column of Fig.~\ref{fig:ttv_data2}); otherwise $\Phi_{\rm TTV} > 0$ (right column).

The TTV phase can be interpreted geometrically. Figure \ref{fig:cartoon} (left panel) illustrates that $\Phi_{\rm TTV} = 0$ corresponds to a transiting conjunction where the inner planet is at periapse (``peri'') while the outer planet is at apoapse (``apo''). The symmetry of this orbital configuration about both the line of conjunctions and the line of sight leads to a TTV that is momentarily zero. 
This symmetry is broken for $\Phi_{\rm TTV} \neq 0$. For the case of non-zero phase, during the transiting conjunction, the inner planet is displaced from its peri, and the outer planet is displaced from its apo (right panel of Fig.~\ref{fig:cartoon}).

Interpreted geometrically as such, $\Phi_{\rm TTV}$ has physical significance. The symmetric peri/apo configuration at conjunction (leading to $\Phi_{\rm TTV}=0$) represents a fixed point of a first-order resonance \citep[e.g.][]{peale_1986}. As we mentioned earlier, dissipation leading to resonance capture drives the system toward this fixed point at $\Delta > 0$. Thus the various dissipative mechanisms --- disc migration, disc eccentricity damping, tides --- that reproduce the observed period ratio statistics (Fig.~\ref{fig:period_ratio_distributions}) predict $\Phi_{\rm TTV}=0$.\footnote{\cite{lithwick_etal_2012} state that their TTV theory does not cover the case where the system is locked in resonance and the resonant argument is librating (see their Appendix A.2 where they state their assumptions). But we have verified that the theory can still be applied to librating systems; in particular a zero-libration (completely locked) system has $\Phi_{\rm TTV} = 0$. The theory does break down if the libration frequency does not scale as $n\Delta$, where $n$ is the mean motion, in which case TTVs are not simple sinusoids. Breakdown occurs for $\Delta \lesssim \mu^{2/3}$, where $\mu$ is the planet-to-star mass ratio (see equation 10 of \citealt{goldreich_1965}).}

Is this expectation of zero TTV phase confirmed by the observations? The answer is no --- Figure \ref{fig:data_3pan}a plots $\Phi_{\rm TTV,1}$ and $\Phi_{\rm TTV,2}$ for a sample of sub-Neptune pairs near 2:1 and 3:2 resonances. Histograms of these phases are plotted in Figure \ref{fig:data_3pan}b, which does not bother to distinguish $\Phi_{\rm TTV,1}$ from $\Phi_{\rm TTV,2}$ since Fig.~\ref{fig:data_3pan}a shows they are nearly equal. The observed distribution of $\Phi_{\rm TTV}$ runs the gamut from -180$^{\circ}$ to +180$^{\circ}$, in apparent contradiction to scenarios of dissipative, migration-induced resonance capture that predict $\Phi_{\rm TTV}=0$. This behavior does not seem to change with distance from the host star, as shown in Figure \ref{fig:data_3pan}c.

Our goal in this paper is to resolve this discrepancy --- to find a formation history for sub-Neptunes that can explain both their period ratio statistics (Fig.~\ref{fig:period_ratio_distributions}) and their non-zero TTV phases (Fig.~\ref{fig:data_3pan}). We will continue to adhere to the picture in which a residual protoplanetary disc drives a pair of sub-Neptunes into resonance while damping their eccentricities \citep{choksi_chiang_2020,macdonald_etal_2020}. To this scenario we will add a third, non-resonant planet. The extra body might be another sub-Neptune. Or the extra body could be a giant planet --- a ``cold Jupiter'' or ``sub-Saturn'' --- 
that radial velocity surveys have found in $\sim$40\% of systems hosting a
transiting sub-Neptune \citep{bryan_etal_2016, zhu_wu_2018, mills_etal_2019a, rosenthal_etal_2021}.

We will show that a third, non-resonant planet can excite the
TTV phase of a resonant pair without interfering with $P_2/P_1$. The excited phase is not constant, but varies as the third body drives secular changes in the pair over a timescale
$P_{\rm sec} \gtrsim P_1/\mu_3 \sim (10^{3} - 10^{5})P_1$, where $\mu_3 = m_3/\Mstar$ is the third planet's mass ratio with the star. This secular time is much longer than the TTV cycle super-period, $P_{\rm super} \sim 10^2P_1$. Over a typical observing window spanning just a few cycles, the TTVs of a resonant pair masquerade as a standard two-planet sinusoid with constant phase.

This paper is organized as follows. Section \ref{sec:sec2} studies a representative three-planet system. Section \ref{sec:params} explores how TTV phases scale with the third planet's mass, eccentricity, and orbital period. Section \ref{sec:summary} summarizes and discusses. Technical details including the equations solved in this paper are contained in the Appendices.

\section{Resonant dynamics including a non-resonant companion}
\label{sec:sec2}
We consider a co-planar pair of sub-Neptunes (labeled ``1'' for the inner body and ``2'' for the outer) that start 
outside the 2:1 mean-motion resonance.
Exterior to this pair orbits another co-planar planet
(``3''), situated far from any first-order resonance with the pair. At the start of our calculations we apply dissipative forces to the interior pair to smoothly damp their eccentricities and have their semi-major axes converge. Once the pair captures into resonance and equilibrates, we shut off these external forces and continue evolving all three planets without dissipation.

We numerically integrate Lagrange's equations (\ref{eqn:n1dot})--(\ref{eqn:w3dot}) for the mean motions $n(t)$, eccentricities $e(t)$, and longitudes of periapse $\varpi(t)$ of the three planets. We feed these numerical solutions into the analytic formulae of \cite{lithwick_etal_2012} --- their equations (1)--(13) --- to compute the TTVs of planets 1 and 2, including TTV phases. Details are provided in Appendix \ref{sec:appendix_A}, which also checks our semi-analytic approach against an $N$-body simulation.

\subsection{Results for period ratios and TTV phases}

\begin{figure*}
\includegraphics[scale=0.30]{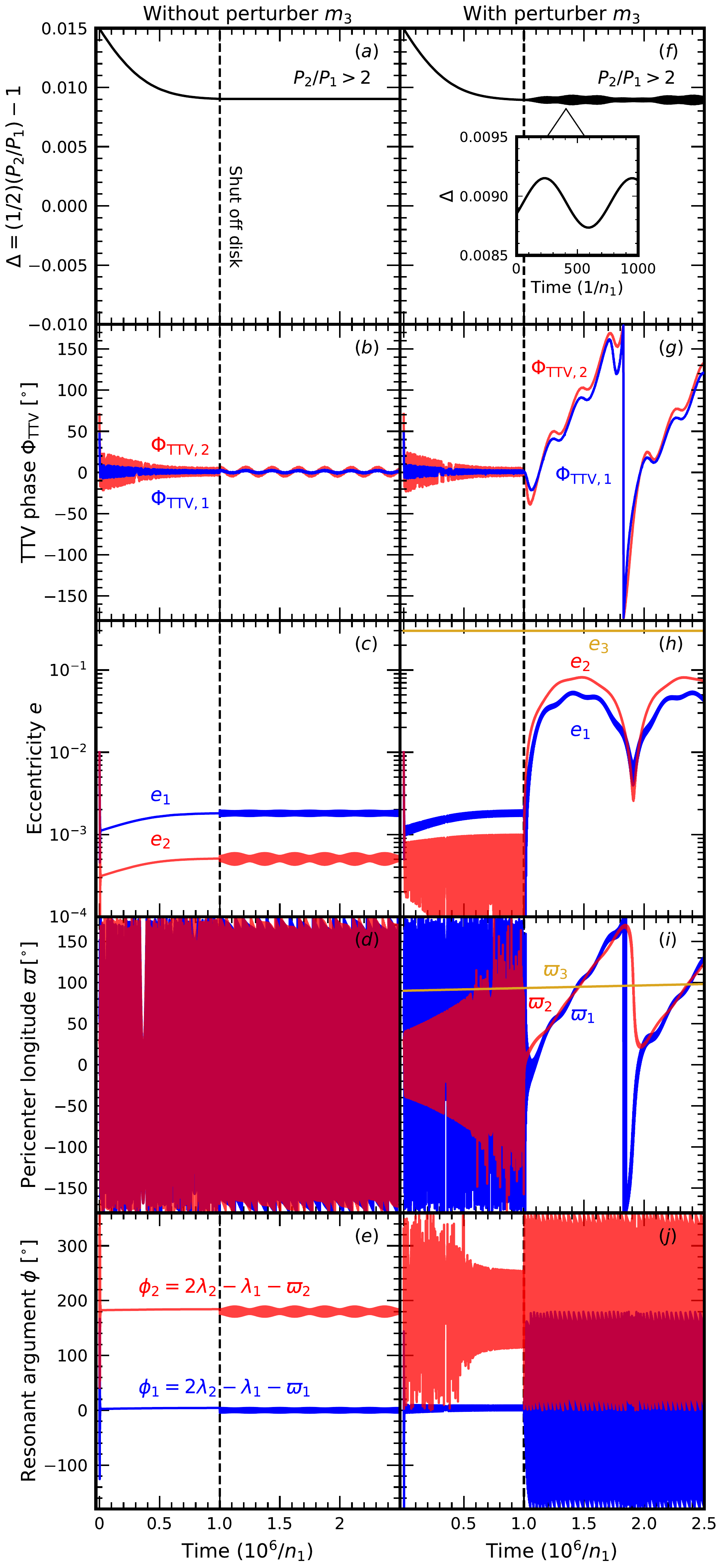} \\
\vspace{-0.25cm}
\caption{Convergent migration of two masses $m_1 = m_2 = 7.5 M_{\oplus}$ into 2:1 resonance, with and without a third non-resonant body. \textit{Left:} Evolution without a third body. The planets capture into resonance with $\phi_1$ and $\phi_2$ locked about 0 and $\pi$, respectively. Eccentricity damping drives the system to a fixed point where $\Phit = 0$. After this equilibrium is reached, the dissipative forces which induce migration and eccentricity damping are shut off (vertical dashed line). \textit{Right:} Same as left, but now with a third body of mass $m_3 = M_{\rm J}$, eccentricity $e_3 = 0.3$, and period $P_3 = 20P_1$. After disc torques are shut off, the third body secularly excites eccentricities and by extension TTV phases. It also forces the pair to apsidally align and precess about its pericentre, breaking the resonant locks while hardly affecting the period ratio.}
\label{fig:rep}
\end{figure*}

\begin{figure}
\centering
\includegraphics[width=0.98\columnwidth]{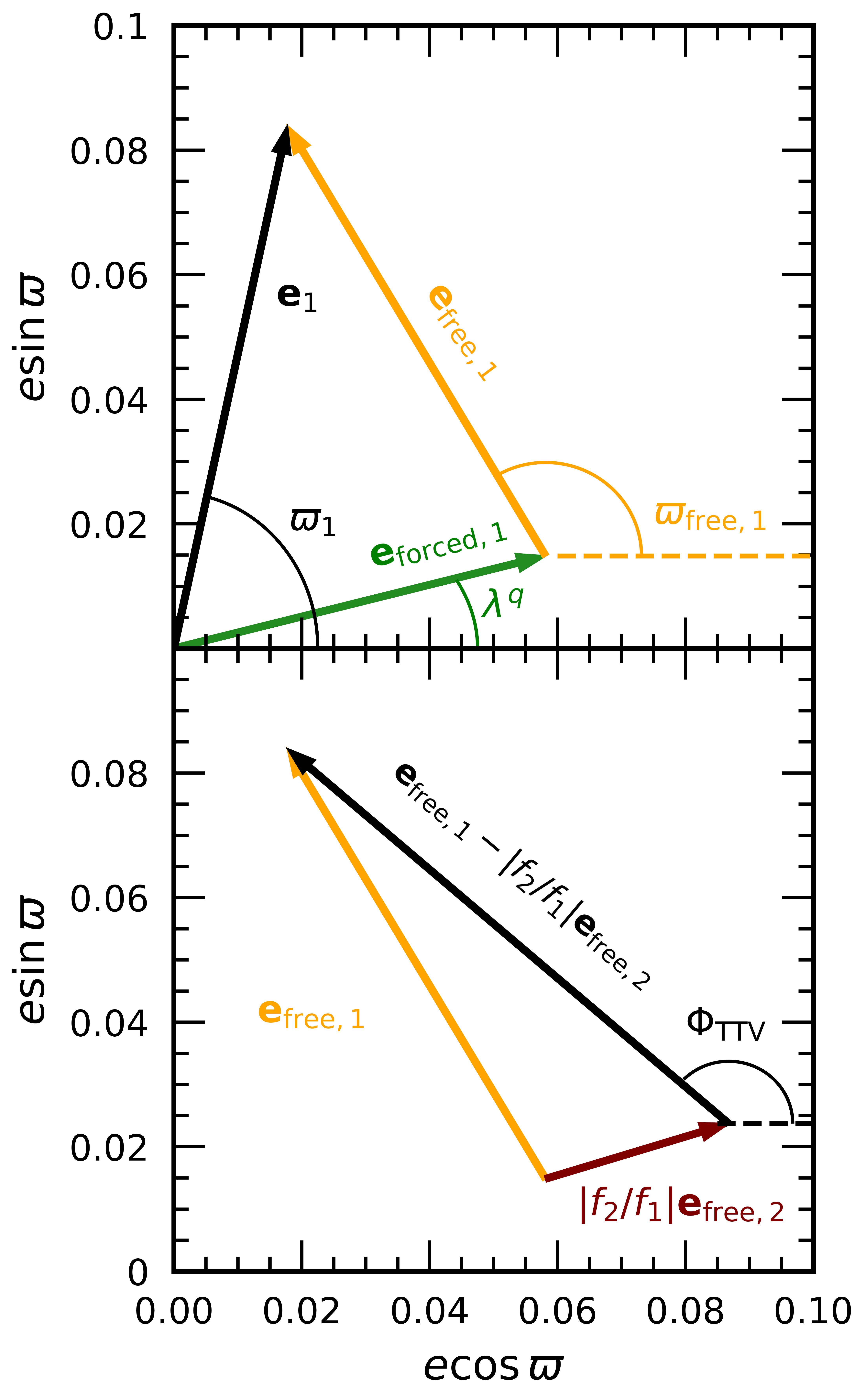} \\
\hspace{-0.5cm}
\caption{\textit{Top:} Decomposition of a planet's eccentricity vector $\mathbf{e_1} = (e_1 \cos \varpi_1, e_1 \sin \varpi_1)$ into forced and free components near a mean-motion resonance. The resonantly forced component has magnitude $e_{\rm forced,1} \sim \mu_2/\Delta$, where $\mu_2$ is the ratio of the resonant companion mass to the star and $\Delta$ is the fractional offset from commensurability. The forced vector points in the direction of the line of conjunctions whose longitude $\lambda^q$ rotates through 2$\pi$ each TTV super-period. By contrast, the free vector is a kind of integration constant that depends on initial or boundary conditions; in this paper, we generate free eccentricities using secular perturbations from a third non-resonant body. Our numerical integrations provide total eccentricity vectors from which we can compute the free contribution $\mathbf{e_{\rm free}} = \mathbf{e} - \mathbf{e_{\rm forced}}$. \textit{Bottom:} How free eccentricity vectors relate to the TTV's phase shift $\Phit$. The phase is controlled not by either $\mathbf{e}_{\rm free}$ vector individually, but by the difference $\mathbf{e}_{\rm free,1} - |f_2/f_1|\mathbf{e}_{\rm free,2}$ (black arrow). When this difference vector has a magnitude that is $\gg \Delta$, its orientation yields $\Phit$ as shown here. If instead its magnitude is $\ll \Delta$, the TTV phase would be $\ll 1$ no matter which way the free eccentricity vectors pointed. 
}
  \label{fig:evec}
\end{figure}

\begin{figure}
\centering
\hspace{-1cm}
\includegraphics[width=0.98\columnwidth]{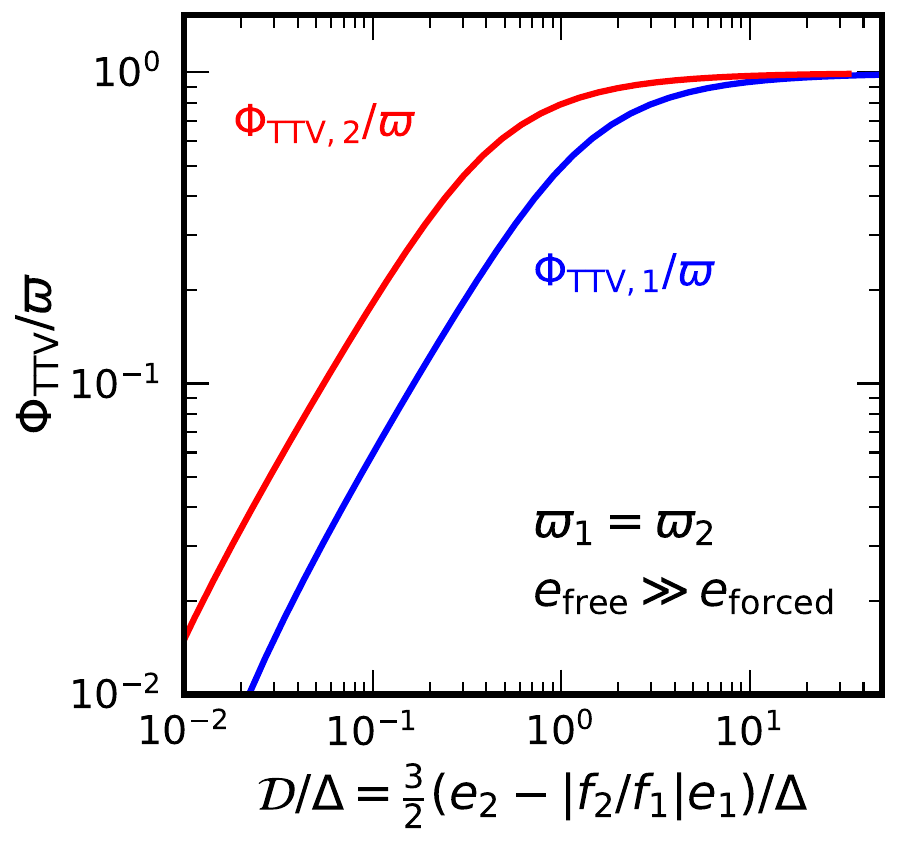} 
\hspace{-0.5cm}
\caption{The quantity $\mathcal{D}/\Delta = 3|e_2-|f_2/f_1|e_1|/(2\Delta)$ serves as a proxy for the TTV phases when eccentricities are dominated by their free components and apses are aligned. These conditions can be satisfied for a near-resonant pair of planets secularly excited by an eccentric third body. The plotted curves give the phases $\Phi_{\rm TTV,1}$ and $\Phi_{\rm TTV,2}$ for planets near 2:1 resonance, derived from equations (\ref{eqn:dt_full})--(\ref{eqn:Z}) assuming $e\simeq e_{\rm free} \gg e_{\rm forced}$ and $\varpi_1=\varpi_2$. When $\mathcal{D}/\Delta \gtrsim 1$, $\Phi_{\rm TTV}$ tracks $\varpi$; otherwise $\Phi_{\rm TTV} \ll 1$ irrespective of $\varpi$.
}
\label{fig:phase_w}
\end{figure}

\begin{figure}
\includegraphics[width=\columnwidth]{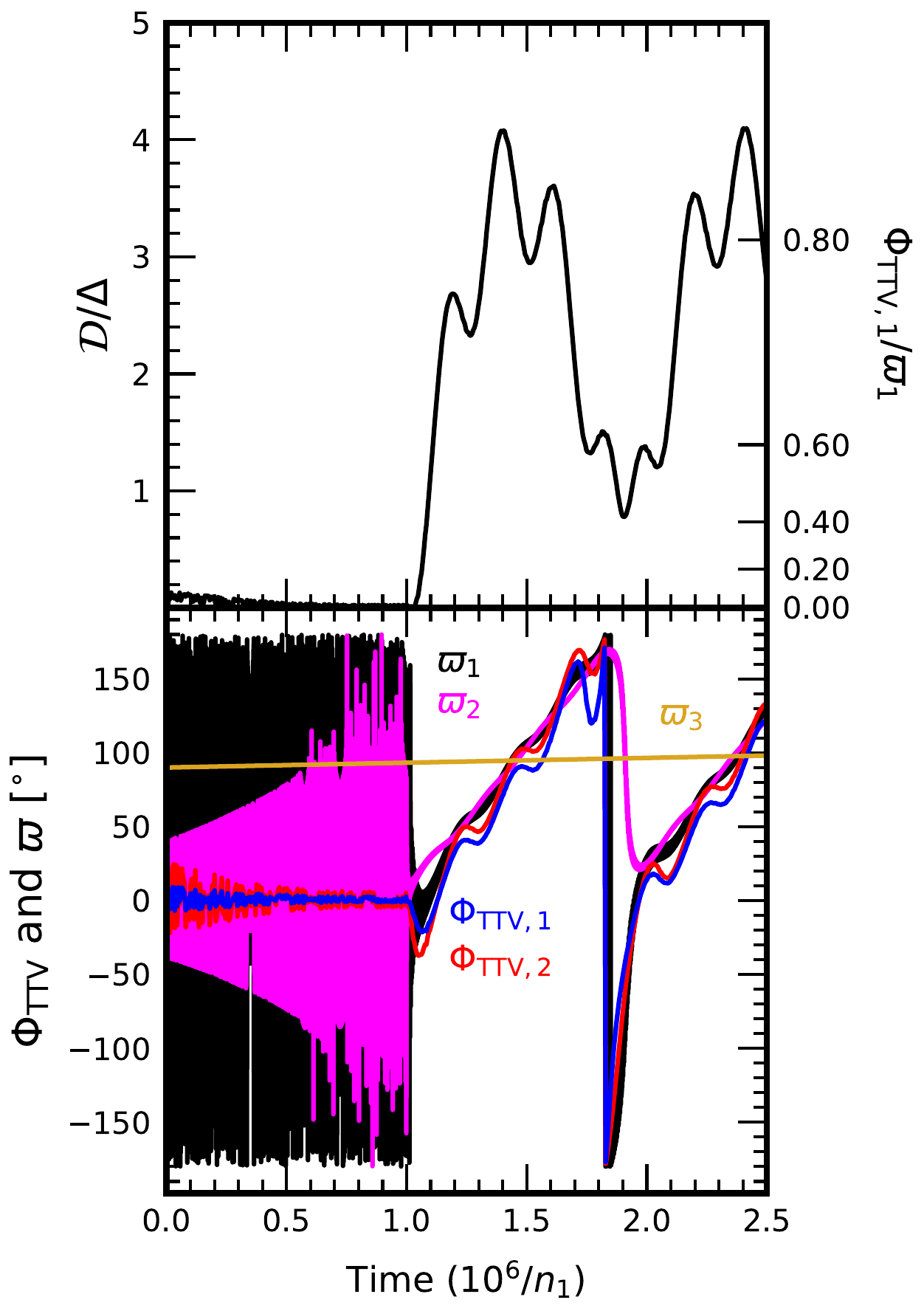} \\ 
\hspace{-0.5cm}
\caption{Demonstration that $\mathcal{D}/\Delta$ is a good proxy for $\Phi_{\rm TTV}$, for the sample evolution in Fig.~\ref{fig:rep} of two near-resonant planets secularly perturbed by a third body. 
When $\mathcal{D}/\Delta \gtrsim 1$ --- a condition that holds after disc torques are removed and the resonant pair's eccentricites are free to be excited by the third body --- the TTV phases track the pair's common periapse longitude $\varpi_1 \simeq \varpi_2$, which varies about the third body's apse. 
}
\label{fig:D_Delta}
\end{figure}

The left column of Figure \ref{fig:rep} reviews how the inner pair evolves in the absence of the third body \citep[e.g.][]{lee_peale_2002, choksi_chiang_2020}. Two sub-Neptunes with masses 
$m_1 = m_2 = 7.5\,\Mearth$ convergently migrate toward the 2:1 resonance. They are captured into resonance, with both resonant arguments $\phi_1$ and $\phi_2$ librating (panel e), and $\Delta$ equilibrating to  
+0.9\% (panel a). The TTV phases rapidly damp to $\lesssim$ 10$^\circ$ from the applied dissipation (panel b). The eccentricities level out as the imposed damping balances excitation by resonant migration (panel c). After the dissipation shuts off,
the eccentricities oscillate a bit more, but overall the system properties do not change much.

The right column of Figure \ref{fig:rep} shows how the same pair evolves when accompanied by a third planet of mass $m_3 = 1 M_{\rm J}$, period $P_3 = 20P_1$, and eccentricity $e_3 = 0.3$. While disc torques are still being applied, the evolution of planets 1 and 2 is not too different from the no-3 case. But after disc torques are removed, the resonant pair is free to respond in full to planet 3. Eccentricities $e_1$ and $e_2$ are secularly driven by planet 3 (panel h), which also forces apsidal alignment, $\varpi_1 \simeq \varpi_2$ (panel i). The resonant locks are broken (panel j). Despite all this, the period ratio $P_2/P_1$ is left at the same equilibrium value as in the two-planet case; compare panels a and f.
The main effect of the third planet on $\Delta$ is to induce modest, 
$\sim$5\% oscillations. The oscillations have both a short period equal to the TTV super-period $P_{\rm super} \approx 500 n_1^{-1}$ (panel f inset) and a long period corresponding to the secular forcing period 
$P_{\rm sec} \approx 10^6\,n_1^{-1} \approx 2 \times 10^3 P_{\rm super}$. It is not surprising that $\Delta$ is relatively unaffected by planet 3, since secular interactions do not change semi-major axes.

By contrast to $\Delta$ which is modulated only mildly by the third body, the TTV phases are dramatically affected. Over $P_{\rm sec}$, both phases $\Phi_{\rm TTV,1}$ and $\Phi_{\rm TTV,2}$ are freed from zero and cycle between -180$^\circ$ and 180$^{\circ}$ (panel g).

\subsection{Analysis of TTV phases} \label{subsec:dissect}
The TTV phases can be dissected using the analytic formulae derived by \citet{lithwick_etal_2012} for a pair of planets 
near a first-order commensurability:\footnote{\cite{lithwick_etal_2012} express the TTVs in terms of complex eccentricities $z = e \exp(i\varpi)$. We have re-written their equations in purely real form. The coefficients $f_1$ and $f_2$ here are called $f$ and $g$ in their paper.}
\begin{align}
&\mathrm{TTV}_1(t)  = 2e_{\rm forced, 1}n_1^{-1}\left[\sin \lambda^q + \frac{3}{2\Delta}\mathcal{Z}\right] \label{eqn:dt_full} \\ 
& \mathrm{TTV}_2(t) = 2e_{\rm forced,2}n_2^{-1}\left[\sin (\lambda^q - \pi) - \frac{3}{2\Delta}\left|\frac{f_1}{f_2}\right|\mathcal{Z}\right] \label{eqn:dt_full2} \\ 
& \mathcal{Z}(t) = \efone\sin(\lambda^q - \wfone)   - \left|\frac{f_2}{f_1}\right|\eftwo\sin(\lambda^q - \wftwo).
\label{eqn:Z}
\end{align}
Expressions for the resonantly forced eccentricities $e_{\rm forced,1}$ and $e_{\rm forced,2}$ and the order-unity coefficients $f_1$ and $f_2$ are given in Appendix \ref{sec:appendix_A}; we note here that to order-of-magnitude, $e_{\rm forced, 1} \sim \mu_2/\Delta$ and vice-versa. The angle $\lambda^q = (q+1)\lambda_2 - q\lambda_1$, where $\lambda$ is the mean longitude, measures the longitude of the line of conjunctions between planets 1 and 2. It sweeps at constant rate $d\lambda^q/dt = -\Delta q n_1$. Our convention is that the observer sits along the positive $x-$axis so that $\lambda^q = 0$ during a transiting conjunction (TC). If the forced term proportional to $e_{\rm forced,1} \sin \lambda^q$ in equation (\ref{eqn:dt_full}) were dominant, then $\Phi_{\rm TTV,1}=0$ (since every  $\lambda^q = 0$ TC would give TTV$_1 = 0$). 
This forced portion of the TTV arises because the orbit precesses apsidally, at a rate $d\varpi_1/dt = d\lambda^q/dt = -\Delta q n_1$ when the resonant argument $\phi_1 = \lambda^q - \varpi_1$ is fixed at 0 in resonance lock. Analogous statements apply to the purely forced term $e_{\rm forced,2} \sin (\lambda^q-\pi)$ in equation (\ref{eqn:dt_full2}).

Adding the terms proportional to $\mathcal{Z}$ leads to $\Phi_{\rm TTV} \neq 0$. These terms depend on the free component of each planet's eccentricity, i.e. the portion of the eccentricity beyond the resonant fixed point value. See the top panel of Figure \ref{fig:evec} for how $e_{\rm free}$ and the related angle $\varpi_{\rm free}$ are extracted from the osculating $e$ and $\varpi$. The free elements $e_{\rm free}$ and $\varpi_{\rm free}$ define a free eccentricity vector $\mathbf{e_{\rm free}}$ which is constant in time; it is a kind of integration constant determined by initial conditions. The forced eccentricity vector's magnitude is also constant, but its direction varies with time at rate $d\lambda^q/dt$.
Thus if the free eccentricity is non-zero, then the osculating eccentricity, given by the vector sum of the forced and free eccentricities, oscillates about the forced (fixed point) value (see also \citealt{lithwick_etal_2012}, their figure 1). Such eccentricity variations drive mean-motion variations (via the Brouwer integrals associated with the resonance; equation A18 of \citealt{lithwick_etal_2012}) which in turn contribute to the TTV.

These mean-motion TTVs due to free eccentricities are generally phase-shifted relative to the aforementioned apsidal precession TTVs. The bottom panel of Fig.~\ref{fig:evec} illustrates this phase shift geometrically. From equation (\ref{eqn:Z}), we construct the difference vector $\mathbf{e_{\rm free,1}} - |f_2/f_1| \mathbf{e_{\rm free,2}}$. The direction of this vector gives $\Phit$, in the limit that the magnitude of this vector $\gg \Delta$ so that the mean-motion TTVs dominate.

Returning to our disc-driven capture scenario, after the disc clears, 
the third body perturbs the inner pair off the fixed point, secularly exciting free eccentricities (panel h of Fig.~\ref{fig:rep}), which
lead to mean motion changes 
(panel f), and
by extension non-zero TTV phases (panel g). 
When total eccentricities $e_1$ and $e_2$ are excited
up to several percent --- an order
of magnitude larger than
their forced values (panel c) --- they are dominated by their
free components, i.e.~$e_{\rm free} \simeq e$ and $\varpi_{\rm free} \simeq \varpi$. In addition, the two resonant planets apsidally align under the influence of the third body, with $\varpi_1\simeq \varpi_2$ varying 
about $\varpi_3$ over a secular timescale (panel i; see also \citealt{beauge_etal_2006} and \citealt{laune_etal_2022}).\footnote{Not only do the standard resonant arguments $\phi_1$ and $\phi_2$ circulate post-disc, but the resonant argument of \cite{laune_etal_2022} (their equation 39) also circulates (data not shown).} 

The conditions $e_{\rm free} \gg e_{\rm forced}$ and $\varpi_1 = \varpi_2$ inserted into equations (\ref{eqn:dt_full})-(\ref{eqn:Z}) imply that each planet's TTV phase lies between 0 and $\varpi$. Where the phase lands in this range is controlled by the ratio
\begin{equation}
    \frac{\mathcal{D}}{\Delta} = \frac{3}{2}\frac{\left| e_1 - \left|{f_2}/{f_1} \right|e_2 \right|}{\Delta} \,,
    \label{eqn:D}
\end{equation}
obtained by comparing the magnitudes of the terms in the square brackets in equations (\ref{eqn:dt_full})--(\ref{eqn:dt_full2}), i.e. the magnitude of the phase-shifted mean-motion TTV to that of the zero-phase precessional TTV. As Figure \ref{fig:phase_w} shows, when $\mathcal{D}/\Delta \ll 1$, both planets have zero TTV phase, whereas when $\mathcal{D}/\Delta \gg 1$, $\Phi_{\rm TTV} = \varpi$. For our sample evolution, $\mathcal{D}/\Delta$ stays above unity over the secular cycle (Figure \ref{fig:D_Delta}a), and the TTV phases approximately track $\varpi$ (Figure \ref{fig:D_Delta}b).

Externally amplifying the 
eccentricities of a resonant
pair of planets
excites $\Phi_{\rm TTV}$ 
more readily than it changes
$\Delta$. Whereas $\Phi_{\rm TTV}/\varpi$ is of order $\mathcal{D}/\Delta$ (for $\mathcal{D}/\Delta \lesssim 1$),
the fractional change in $\Delta$
is of order $\sim$$10e_{\rm forced} \mathcal{D}/\Delta$, by virtue of the
Brouwer constants of the motion associated
with the resonance (equation A19 of \citealt{lithwick_etal_2012}). 
The relative insensitivity of $\Delta$
means that the fits obtained to observed period ratios 
within the scenario
of disc-driven dissipation and migration
\citep{choksi_chiang_2020} should not change
much when eccentricity forcing
by an external companion is added.

\section{Parameter dependence}
\label{sec:params}

\begin{figure*}
\centering
\hspace{-1cm}
\includegraphics[width=0.98\textwidth]{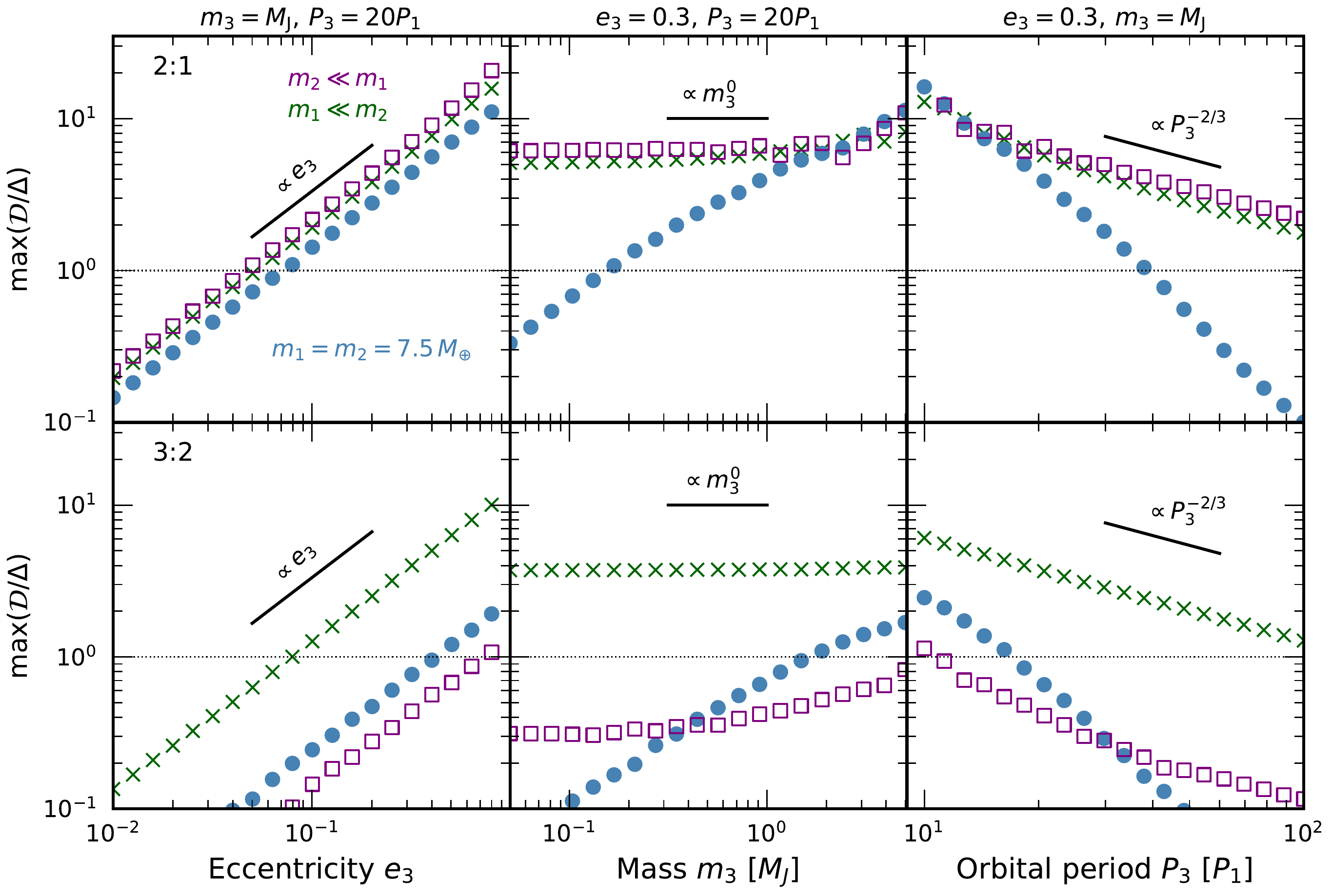} \\ 
\hspace{-0.5cm}
\caption{How $\mdd$, our proxy for TTV phase, varies with the properties of an eccentric third body located exterior to a resonant pair of sub-Neptunes. These results are obtained from  the same procedure of initially imposing convergent migration and eccentricity damping on the resonant pair (section \ref{sec:sec2}). In each column we vary one property of the third body while keeping the others held fixed at values listed at the top of the figure. For TTV phases to be non-zero, we need $\mdd \gtrsim 1$, which is possible for sufficiently eccentric and massive third bodies situated close enough to the resonant pair (but not so close as to become mean-motion resonant with the pair). It is easier to reach $\mdd \gtrsim 1$ when the resonant planets are less  coupled to each other, so that the third body can more strongly interfere with their behavior; thus $\mdd \gtrsim 1$ tends to obtain more readily for the 2:1 resonance than for the more closely spaced 3:2, and when one member of the resonant pair is much less massive than the other (especially when $m_1 \ll m_2$; see also Appendix \ref{sec:appendix_B}). 
Power-law slopes expected in the test particle limits $m_1 \ll m_2$ and $m_2 \ll m_1$ are indicated. }
  \label{fig:parameter_survey}
\end{figure*}

\begin{figure*}
\centering
\hspace{-1cm}
\includegraphics[width=0.98\textwidth]{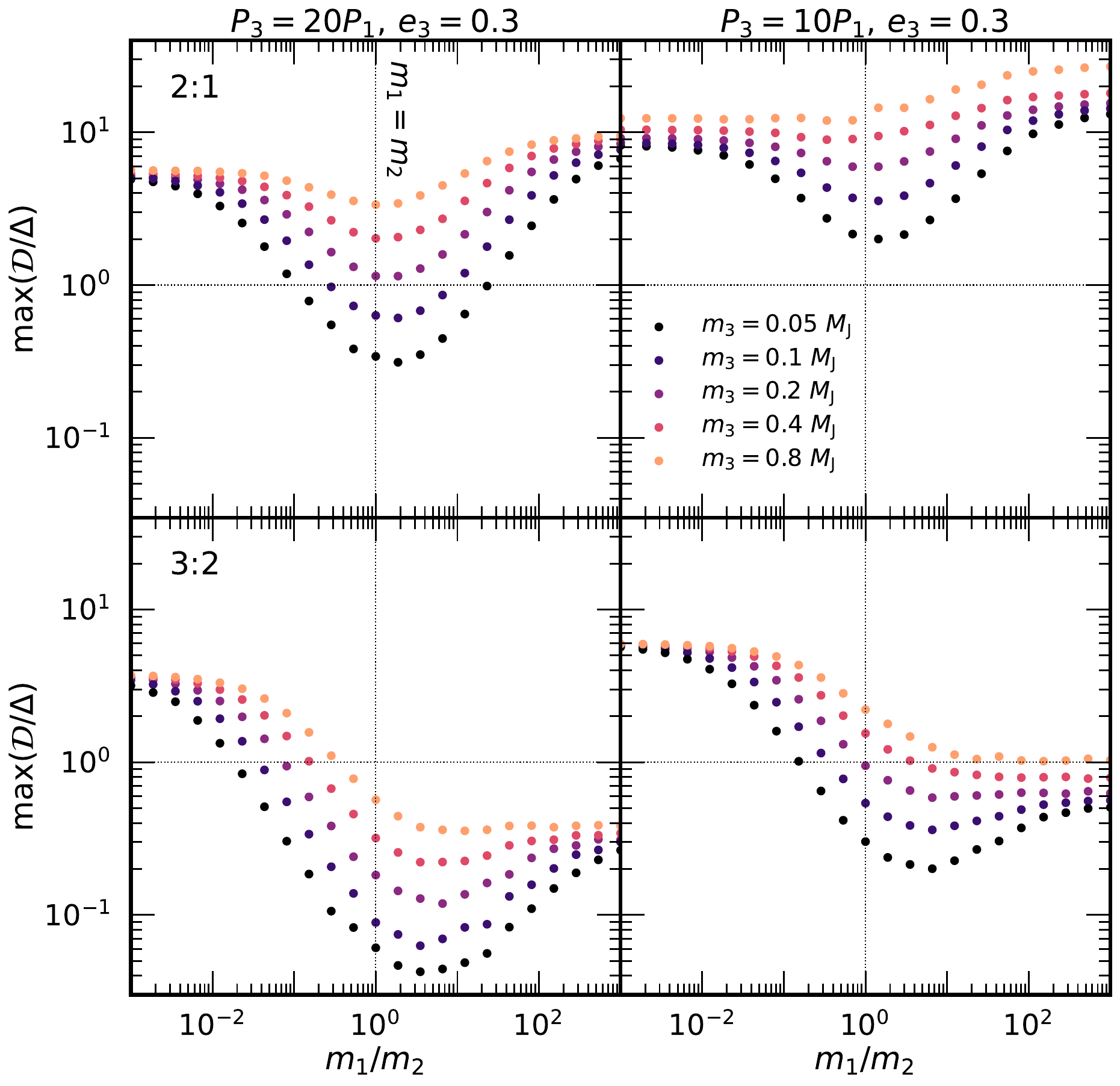} \\ 
\hspace{-0.5cm}
\caption{Similar to Figure \ref{fig:parameter_survey}, but now exploring how our TTV phase proxy, $\mdd$, varies with the mass ratio $m_1/m_2$ of the resonant pair, and with the mass $m_3$ of the external perturber, for $P_3 = 20P_1$ (left column) and $P_3 = 10P_1$ (right column), at fixed $e_3 = 0.3$. For the 2:1, the trends are as expected: $\mdd$ and by implication the average $|\Phit|$ are higher when the third body perturber is closer to the pair and more massive, and when the pair's masses differ substantially (test particle regime). These trends are largely shared by the 3:2 with some differences: in addition to an overall lower $\mdd$ (a consequence of the 3:2 resonance being stronger than the 2:1 and therefore more impervious to external secular forcing), the $m_2 \ll m_1$ test particle limit yields lower $\mdd$ than the opposing $m_1 \ll m_2$ limit. These technical differences are explored in Appendix \ref{sec:appendix_B}.}
\label{fig:mratio}
\end{figure*}

\begin{figure*}
\hspace{-1cm}
\includegraphics[width=0.98\textwidth]{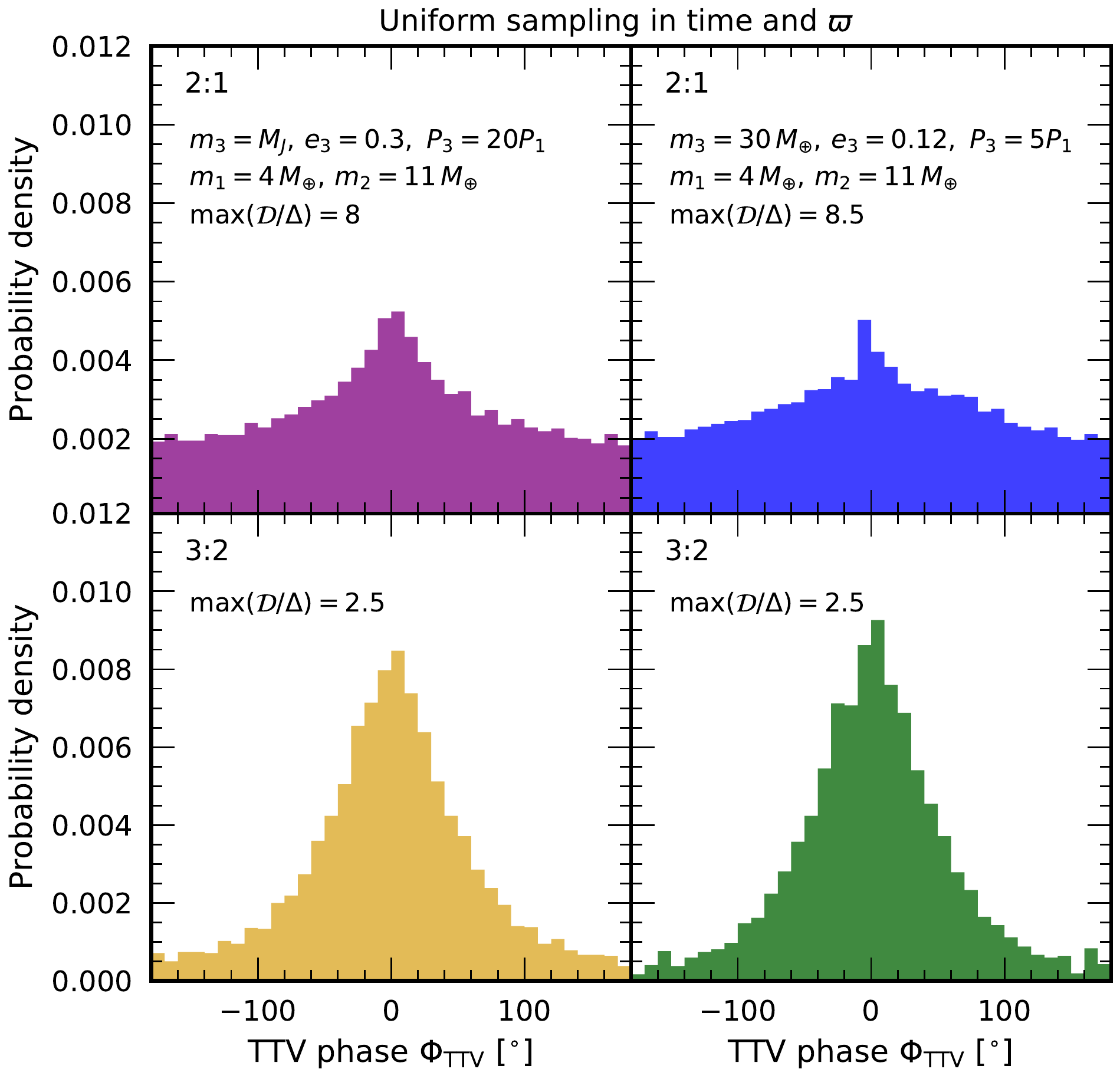} \\ 
\hspace{-0.5cm}
\caption{How $\Phit$ distributes in time for sub-Neptunes in the 2:1 resonance (top panels) and the 3:2 (bottom panels), for parameters listed at the top of each column. Each panel uses 30 simulations of resonant pairs perturbed by an exterior third body whose initial longitude of periastron $\varpi_3$ is drawn uniformly from 0 to $2\pi$; TTV phases are sampled at uniform intervals of time after migration and eccentricity damping of the pair are shut off.  
Between the left and right columns, parameters are chosen to give approximately equal values of $\mdd$ and thus similar distributions of $\Phit$. The more time a system spends with $\dd \gtrsim 1$, the closer the TTV phases track $\varpi_1$ and $\varpi_2$ --- both of which follow $\varpi_3$ --- and the broader the time-sampled distribution of $\Phit$.}
\label{fig:mc_phases}
\end{figure*}
We explore how the non-resonant third planet's eccentricity $e_3$, mass $m_3$, and orbital period $P_3$ affect the TTV phases of the resonant pair. As we saw in the previous section, the larger is $\mathcal{D}/\Delta$ (equation \ref{eqn:D}), the more the TTV phases can deviate from zero. When $\mathcal{D}/\Delta > 1$, the TTV phases approximately track $\varpi_1$ or $\varpi_2$ --- these apsidal longitudes are nearly the same under secular driving by an eccentric third body. The pair's apses in turn follow $\varpi_3$, which is free to take any value between 0 and $2\pi$ relative to our line of sight. Thus as a proxy for the range of accessible TTV phases, we measure how $\mdd$ depends on the properties of the third planet. To build intuition, we first study the test particle limits in which the resonant planets have very different masses,
$m_1 \ll m_2 \lesssim m_3$ or $m_2 \ll m_1 \lesssim m_3$. Then we see how our results change when $m_1$ is comparable to $m_2$,
keeping the total mass fixed to $m_1 + m_2 = 15\,\Mearth$.

Figure \ref{fig:parameter_survey} plots $\mdd$ as measured from our numerical integrations varying one parameter at a time around a fiducial set $\{e_3,\,m_3,\,P_3\} = \{ 0.3,\,1\,\Mj,\,20P_1\}$. In both test particle limits (purple squares and green crosses) near the 2:1 resonance (top row), $\mdd$ spans a factor of $\sim$100 across our explored parameter space. Most of this variation stems from changes in $\mathcal{D}$ (and not $\Delta$; see the end of section  \ref{subsec:dissect}), reflecting how strongly the third body secularly amplifies $e_1$ and $e_2$. Reducing the perturber's eccentricity (left panels of Fig.~\ref{fig:parameter_survey}) or moving it farther from the pair (right panels) weakens this amplification. At the same time, the perturber's mass seems to hardly matter (middle panels); in the test particle limit, perturbers weighing anywhere from 15 $\Mearth$ to a few $M_{\rm J}$ produce nearly the same $\mdd$. Pairs near the 3:2 resonance (bottom row) follow all the same trends, but their $\mdd$ values are systematically lower than for the 2:1. 
The TTV phases of 3:2 pairs are harder to excite by an external perturber than those of 2:1 pairs, a consequence of 3:2 pairs being closer together and enjoying a stronger resonant interaction which better isolates them from external influences. More details about this difference between the 3:2 and 2:1 are given in Appendix \ref{sec:appendix_B}.

Our numerical results in the test particle limits suggest power-law scalings that can be reproduced as follows. When $m_1 \ll m_2$, the eccentricity $e_2$ is forced by $m_3$ and not $m_1$. If the perturber is distant ($P_3 \gg P_2$) and massive ($m_3 \gg m_2$), it secularly excites $e_2$ up to $\mathrm{max}(e_2) = (5/2)(P_2/P_3)^{2/3} e_3$ \citep{murray_dermott_1999}. If we further assume that $\Delta$ is constant and that $e_1$ tracks $e_2$, then
\begin{align}
   \mdd &\propto e_3 m_3^0 P_3^{-2/3} \,,
   \label{eqn:D_symmetric}
\end{align}
scalings which are consistent with our numerical data in the appropriate limits for both the 2:1 and 3:2. The same scalings apply when $m_2 \ll m_1 \ll m_3$.

 \rev{In the more realistic case where $m_1$ and $m_2$ are comparable (blue circles in Fig.~\ref{fig:parameter_survey})}, $\mdd$ and by extension TTV phases drop relative to the $m_1\ll m_2$ test particle case, precipitously in some regions of parameter space. Allowing for $m_1 \sim m_2$ strengthens the mutual interaction of the resonant pair and shields them from perturbations by an external third body --- see also Figure \ref{fig:mratio} which underscores the sensitivity of $\mdd$ to perturber mass when $m_1 \sim m_2$. To keep $\mdd > 1$ for the 3:2 resonance --- which automatically keeps $\mdd > 1$ for the 2:1 --- it suffices to have an eccentric Jupiter-mass perturber at $P_3 \sim 15 P_1$; or an eccentric Saturn-mass perturber with $P_3 \sim 10P_1$; or an eccentric Neptune-mass perturber with $P_3 \sim 5 P_1$.
Interestingly, in some regions of parameter space for the 3:2, $\mdd$ actually increases for $m_1 = m_2$ relative to the $m_2 \ll m_1$ test particle case (Fig.~\ref{fig:parameter_survey}), but the enhancements are less than factors of 2-3.

So far in this section we have used $\mdd$ as a proxy for whether TTV phases can be large or must remain small. 
In reality $\dd$ and by extension $\Phit/\varpi$ vary with time over the secular cycle (Figs.~\ref{fig:rep} and \ref{fig:D_Delta}). 
Time-sampled distributions of $\Phit$ generated from ensembles of 3-planet numerical integrations are plotted in Figure \ref{fig:mc_phases}. For each integration, set up the same way as in section \ref{sec:sec2} and with an initial $\varpi_3$ drawn randomly from a uniform distribution between 0 and $2\pi$ (reflecting the isotropy of space), we record $\Phi_{\rm TTV,1}$ and $\Phi_{\rm TTV,2}$ at evenly spaced time intervals over one secular cycle post-dissipation. Each panel in Fig.~\ref{fig:mc_phases} compiles time samples from 30 such integrations, each with a different starting $\varpi_3$, and combining $\Phi_{\rm TTV,1}$ and $\Phi_{\rm TTV,2}$ because they are similar. The resultant distributions of $\Phi_{\rm TTV}$ concentrate at zero, more strongly for the 3:2 since it is less sensitive to secular forcing by the third body. The system parameters for the left and right panels are chosen to give practically identical $\Phit$ distributions, illustrating the degeneracy between $e_3$, $m_3$, and $P_3$.

\begin{figure*}
\includegraphics[width=0.98\textwidth]{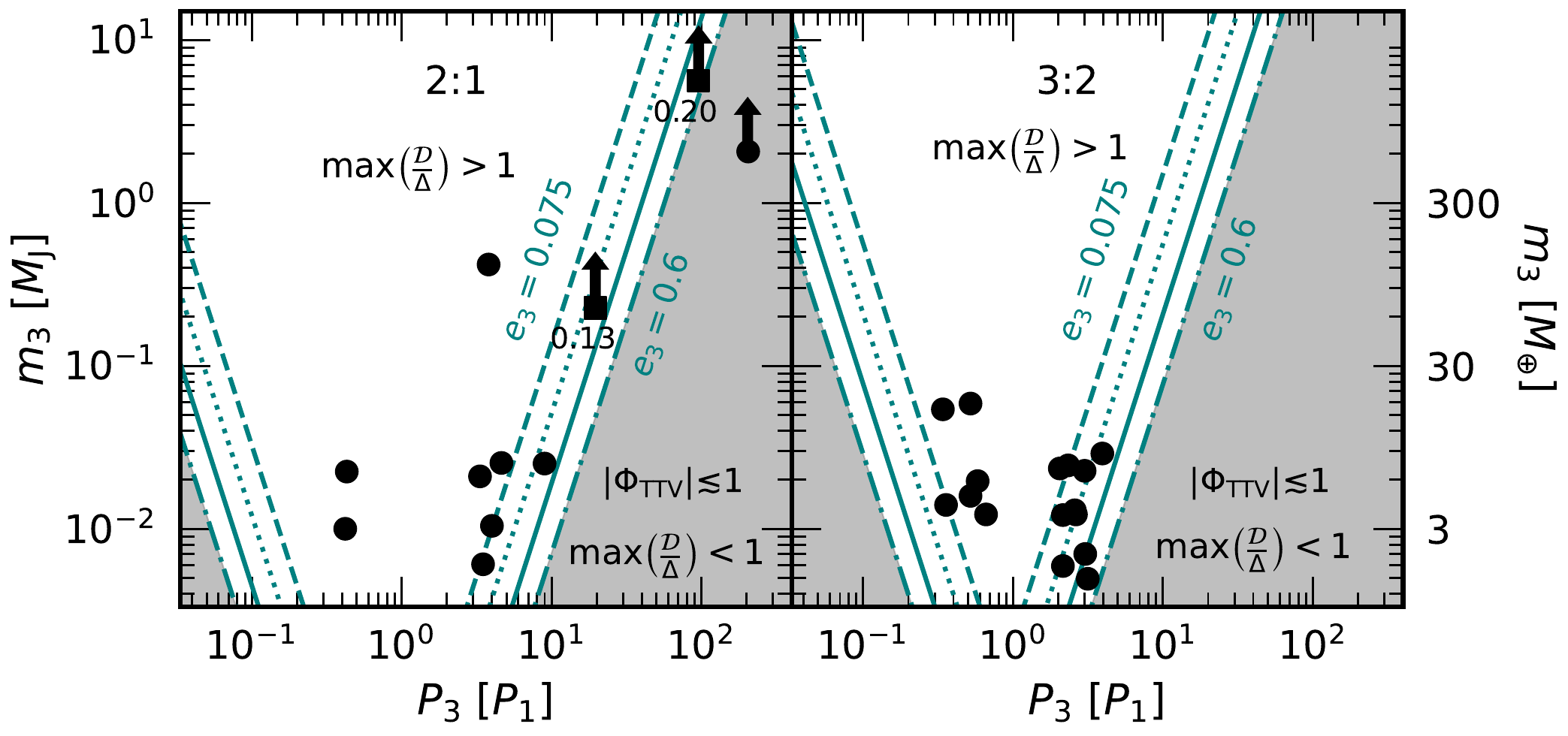} \\ 
\caption{Properties of third-body perturbers needed to secularly excite TTV phases of sub-Neptunes in 2:1 (left) and 3:2 (right) resonance. White regions denote the parameter space where $\mdd > 1$ and TTV phases can be large;  perturbers here need to be massive and close to the resonant pair (either interior or exterior). The third body also needs to be eccentric; minimum values of $e_3$ ranging from 0.075 to 0.6 (in factor of 2 increments) correspond to different pairs of lines enclosing the space where $\mdd > 1$. These lines are computed assuming the resonant sub-Neptunes have equal mass ($7.5 M_\oplus$ each), which tends to yield conservatively low values of $\mdd$ (Fig.~\ref{fig:mratio}).
Symbols denote known third-body perturbers in observed resonant sub-Neptune systems with sinusoidal TTVs and well-defined phases, as drawn from the \protect \cite{rowe_etal_2015} catalog. The masses $m_3$ of these known perturbers are inferred either from a radius-mass relation (figure 12 of \protect \citealt{rogers_owen_2021}) applied to measured transit radii, or from radial velocities which provide lower limits. Black squares are reserved for third bodies with measured $e_3$ as labeled. Some comments on individual cases: the perturber for the 2:1 system plotted on the extreme right cannot excite TTV phases unless its eccentricity and/or mass are large. The perturber with $e_3 = 0.20$ is also incapable, unless its mass is much larger than its minimum. By contrast, the perturber with $e_3 = 0.13$ has the necessary properties, marginally. Other perturbers of lower mass can secularly excite TTV phases provided they are sufficiently eccentric. }
\label{fig:m3p3}
\end{figure*}

\section{Summary and Discussion}
\label{sec:summary}

Many mean-motion resonant sub-Neptunes exhibit sinusoidal transit timing variations (TTVs) whose phases $\Phit$ are non-zero, consistent with circulation or large-amplitude libration about their fixed points. 
This observation is surprising because dissipative processes that capture  pairs into resonance also 
drive them to their fixed points, 
damping $\Phit$ to zero. In this paper we showed how secular forcing of a resonant pair by an eccentric third body can phase-shift TTVs. The third body can force the pair into apsidal alignment, and $\Phit$ can track their common longitude of pericenter. While the non-resonant third body knocks the pair off the resonant fixed point,
it does not much alter their
period ratio, $P_2/P_1$. Thus
scenarios that do alter $P_2/P_1$ to reproduce the observed ``peaks'' and ``troughs'' in the period ratio histogram can be simply augmented by a third-body secular perturber to account for non-zero TTV phases. In particular, 
orbital migration and eccentricity damping in a residual protoplanetary disc can drive a fraction of sub-Neptunes into resonance with $P_2/P_1$ 
just wide of perfect commensurability, as observed \citep{choksi_chiang_2020}; and if such capture plays out in the presence of a sufficiently massive and eccentric non-resonant third body, the TTVs can have non-zero phases after the disc clears, also as observed.

Figure \ref{fig:m3p3} summarizes the parameter space favoring the secular excitation of TTV phases and shows where known companions to TTV pairs lie in this space. Two such companions have had their eccentricities $e_3$ measured from radial velocities (RVs). One of them, a sub-Saturn with $e_3 = 0.13^{+0.13}_{-0.09}$, looks marginally capable of exciting $\Phit$, which is measured to be $\sim$20$^\circ$.
The other, a super-Jupiter
with $e_3 = 0.20^{+0.01}_{-0.01}$, seems too distant to generate the $\sim$$100^{\circ}$ phase of its associated TTV pair. Most of the remaining companions in Fig.~\ref{fig:m3p3} are other transiting sub-Neptunes lacking measured eccentricities. We see that their eccentricities $e_3$ would have to be $\gtrsim 0.05$-0.15 to secularly excite TTV phases. 
How such eccentricities were acquired and maintained against disc damping would need to be explained. 

\rev{Fig.~\ref{fig:m3p3} also shows that many of these smaller-mass third-body perturbers lie near period commensurabilities with their resonant pairs. The perturbers have $\Delta$'s of $\sim$5-20\%, larger than the $\Delta$'s of $\sim$1\% exhibited by the pairs. In preliminary $N$-body integrations of these systems, we have not found near-resonant forcing by the third body to affect the TTV phase of the resonant pair. We do find that when the third body's $\Delta$ $\lesssim$ the pair's $\Delta$, the TTVs of the pair are no longer sinusoidal. Non-sinusoidal TTVs are observed for extensive resonant chains like TRAPPIST-1, \textit{Kepler-60}, and TOI-1136 (e.g. fig. 2 of \citealt{agol_etal_2021}; fig. 10 of \citealt{jontof-hutter_etal_2016}; fig. 5 of \citealt{dai_etal_2023}). Although a phase seems challenging to define for a non-sinusoidal signal, one can always mark when two (or more) planets transit the star at the same time (as we have done in Fig.~\ref{fig:ttv_data2}). Future work could explore how the relative timing of transiting conjunctions in resonant chains depends on formation history, including how much eccentricity and semimajor axis damping each planet experienced.}

Transit timing variations can be combined with transit duration and/or RV data to place tighter dynamical constraints on systems. Radial velocities can pin down planet masses, and transit durations can help measure individual planet eccentricities, breaking the
``mass-eccentricity'' and ``eccentricity-eccentricity'' degeneracies inherent to using TTVs alone \citep{lithwick_etal_2012}. For example, \cite{dawson_etal_2021} combined RV and transit data to measure a 60$^{\circ} \pm 2^\circ$ libration amplitude for the resonant argument $\phi$ of the sub-Neptune TOI-216b in 2:1 resonance with the gas giant TOI-216c. From our study we have learned that a third body can secularly excite resonant libration amplitudes, all the way up to circulation, without materially changing the period ratio at the time of resonance capture. For the resonant planet pairs \textit{Kepler}-90gh and K2-19bc, \cite{liang_etal_2021} and \cite{petigura_etal_2020} measured aligned apsides, which our study has shown is possible if these pairs are secularly forced by third bodies. The apsidal alignment in \textit{Kepler}-90 might be enforced by any of the six other transiting planets in the system. For K2-19, we find that an eccentric $30\,\Mearth$ planet orbiting with $P_3 = 5P_1 \approx 40$ days could explain the observed alignment. With an RV semi-amplitude of 
7 m/s, such a body could hide below the current noise threshold
of 13 m/s. Continued RV monitoring will decide whether an eccentric perturber exists, or if the observed alignment requires another explanation such as eccentricity pumping by the natal disc \citep{laune_etal_2022}.

\cite{wu_lithwick_2013} and \cite{hadden_lithwick_2014} used the observed distribution of TTV phases to estimate the free eccentricities of resonant sub-Neptunes, assuming that the planets in a given resonant pair have uncorrelated apsidal orientations and eccentricities. This assumption does not hold in our scenario where both planets are secularly forced by a third body into apsidal alignment and to have a particular ratio of eccentricities. Whereas they inferred free eccentricities of 1-2\%, we would derive, in the context of our model, free eccentricities as large as the planets' total eccentricities, which  can be secularly forced by an eccentric third body up to values of $\sim$0.2. Do severely underestimated free eccentricities affect the distribution of planet masses they derived using TTV amplitudes --- a distribution known to be consistent with independent mass measurements from RVs? Fortunately, no --- the TTV amplitudes depend on free eccentricities only through the particular linear combination of free eccentricity vectors shown in the bottom panel of Fig.~\ref{fig:evec}, and this vector combination is reliably measured from the observed TTV phase.

\rev{We have assumed in this work that gas disc torques always drive planet pairs to their resonant fixed points. If the disc is turbulent, density fluctuations add a random walk component to a planet's orbital evolution \citep[][]{laughlin_etal_2004, adams_etal_2008, rein_papaloizou_2009, batygin_adams_2017}. Planetesimal discs also drive migration having both smooth and stochastic contributions (\citealt{murray-clay_chiang_2006}; \citealt{ormel_etal_2012}; \citealt{nesvorny_voh_2016}).
On the one hand stochasticity can excite free eccentricities and TTV phases \citep{goldberg_batygin_2022}. But too much stochasticity can wipe out the peak-trough asymmetries seen in the {\it Kepler} period ratios (Fig.~\ref{fig:period_ratio_distributions}; \citealt{lissauer_etal_2011, fabrycky_etal_2014, steffen_hwang_2015, choksi_chiang_2020}). \citet{rein_2012} argues that a suitably tuned mix of stochastic and smooth migration torques can reproduce the observed period ratio distributions. This study's reproduction of resonant peak-trough structures would be better assessed by replacing their cumulative distributions with differential ones --- the modeled troughs short of commensurability might actually be too filled in. What the stochasticity and migration parameters advocated by \citet{rein_2012} predict for TTVs and in particular TTV phases is unknown; it is also not clear that the implied period distribution (distinct from the period ratio distribution) satisfies observations (cf.~\citealt{elee_chiang_2017}). \citet{goldberg_batygin_2022} neglect the problem of the peak-trough asymmetries in the period ratios, as their plots record $|\Delta|$ instead of $\Delta$.}

\section*{Acknowledgements}
We are indebted to Yoram Lithwick for helping to launch this work. We also thank Eric Agol, Konstantin Batygin, Lister Chen, Rebekah Dawson, Dan Fabrycky, Eric Ford, Max Goldberg, Sam Hadden, Matt Holman, Shuo Huang, Sarah Millholland, Erik Petigura, Hanno Rein, Jason Rowe, Andrew Vanderburg, Shreyas Vissapragada, and Jack Wisdom for useful exchanges. \rev{An anonymous referee provided a constructive report.} Simulations were run on the Savio cluster provided by the Berkeley Research Computing program at the University of California,
Berkeley (supported by the UC Berkeley Chancellor, Vice
Chancellor for Research, and Chief Information Officer). This research has made use of the NASA Exoplanet Archive, which is operated by the California Institute of Technology, under contract with the National Aeronautics and Space Administration under the Exoplanet Exploration Program. Financial support was provided by NSF AST grant 2205500, and an NSF Graduate Research Fellowship awarded to NC.

\section*{Data availability}
No new data were collected as part of this work.




\bibliographystyle{mnras}
\bibliography{planets_nick} 



\appendix

\section{Equations solved}
\label{sec:appendix_A}
The results in this paper were obtained by integrating Lagrange's equations of motion, using a disturbing function that
includes resonant and secular terms to leading order. We list these equations in section \ref{subsec:eom} and check their validity in section \ref{subsec:assumptions}.

\subsection{Equations of motion}
\label{subsec:eom}
We consider two planets lying near a $(q+1)$:$q$ mean-motion resonance (where $q$ is a positive integer) and accompanied by a third non-resonant planet. The equations below are for the case where the third planet orbits exterior to the near-resonant pair. We also modeled third bodies lying interior to the pair, but do not list the equations for that case; they may be derived in a straightforward way from the exterior case \citep{murray_dermott_1999}.

The inner resonant planet evolves according to:
\begin{align}
   \dot{n}_1 = & 
   \, \, 3qn_1^2\alpha_{12}\mu_2\left(e_1 |f_1|\sin\phi_1 - e_2|f_2|\sin\phi_2\right) \nonumber \\
   &+ \frac{3n_1}{2t_{a}} + \frac{pn_1e_1^2}{t_{e}}  \label{eqn:n1dot} \\
    \dot{e}_1 = & \,\,n_1 \alpha_{12} \mu_2 |f_1| \sin \phi_1   \nonumber \\ 
    &- n_1\alpha_{12}\mu_2|\CB(\alpha_{12})|e_2\sin(\varpi_1 - \varpi_2)   \nonumber \\ 
    &- n_1\alpha_{13}\mu_3|\CB(\alpha_{13})|e_3 \sin (\varpi_1 - \varpi_3)  \nonumber \\   
    &-\frac{e_1}{t_{e}} 
    \label{eqn:e1dot} \\ 
    \dot{\varpi}_1 = & -\frac{|f_1|\alpha_{12}n_1\mu_2}{e_1}\cos\phi_1  \nonumber  \\ 
    &+  n_1\alpha_{12}\mu_2\left[2|\CA(\alpha_{12})| - |\CB(\alpha_{12})|\frac{e_2}{e_1}\cos(\varpi_1 - \varpi_2)\right] \nonumber  \\ 
    &+n_1\alpha_{13}\mu_3\left[ 2|\CA(\alpha_{13})| - |\CB(\alpha_{13})|\frac{e_3}{e_1}\cos(\varpi_1 - \varpi_3)\right] \\  
    \dot{\phi}_1 =& \,\,(q+1)n_2 - qn_1 - \dot{\varpi}_1 \label{eqn:phi1dot} 
\end{align}
where $n$ is the mean motion, $e$ is the eccentricity, $\varpi$ is the longitude of periapse, 
$\phi_i = (q+1)\lambda_2 - q\lambda_1 - \varpi_i$ is the resonant argument describing where conjunctions happen in orbital phase (i.e. relative to periapse), $\lambda$ is the mean longitude, $\mu$ is the planet-to-star mass ratio, and subscripts 1, 2, and 3 refer respectively to the inner resonant planet, the outer resonant planet, and the third exterior non-resonant planet. The coefficients $f_1$, $f_2$, $\CA$, and $\CB$ depend on the semimajor axis ratio $\alpha_{ij} = a_i/a_j = \left( P_i / P_j \right)^{2/3}$:
\begin{align}
    f_1 &= -\frac{1}{2}\left[2(q+1) + \alpha\frac{d}{d\alpha}\right]b_{1/2}^{q+1}(\alpha)      \label{eqn:f1} \\
    f_2 &= \frac{1}{2}\left[2q + 1 + \alpha\frac{d}{d\alpha}\right]b_{1/2}^{q}(\alpha) - 2\alpha \delta_{q,1} \label{eqn:f2} \\ 
    \CA &= \frac{1}{8}\left[2\alpha\frac{d}{d\alpha}b_{1/2}^{0}  + \alpha^2\frac{d^2}{d\alpha^2}b_{1/2}^{0}  \right] \label{eqn:C1} \\ 
    \CB &= \frac{1}{4}\left[2b_{1/2}^{1} - 2\alpha\frac{d}{d\alpha}b_{1/2}^{1} - \alpha^2\frac{d^2}{d\alpha^2}b_{1/2}^{1}\right]  \label{eqn:C2} \\ 
    b_{s}^{j}(\alpha) &= \frac{1}{\pi}\int_0^{2\pi}\frac{\cos(j\psi)}{\left(1 - 2\alpha\cos\psi + \alpha^2\right)^{s}}d\psi  \label{eqn:bsj}
\end{align}
\citep{murray_dermott_1999}. 
The term $\delta_{q,1}$ in (\ref{eqn:f2}) is the Kronecker delta and accounts for the contribution of the indirect potential of the 2:1 resonance;
we will return to this term in Appendix \ref{sec:appendix_B}. We keep the coefficients (\ref{eqn:f1})--(\ref{eqn:C2}) fixed at their initial values (Table \ref{tab:coeff}) since they vary negligibly over the course of our integrations.

\begin{table}
\begin{tabular}{|c|c|c|c|c|c|}
\hline 
$P_j$:$P_i$ & $\alpha_{ij}$ & $f_1$ & $f_2$ & \,\,$\CA$ & $\CB$    \\ \hline 
3:2 & 0.76 & -2.02 & 2.48 & 1.15 & -2.00 \\ 
2:1 &  0.63   & -1.19 & 0.43 & 0.39 & -0.58 \\ 
20:1 &  0.14   &   N/A     &  N/A      & 0.0072    & -0.0024 \\
\hline 
\end{tabular}
\caption{Coefficients used in the equations of motion (\ref{eqn:n1dot})-(\ref{eqn:w3dot}) and defined in  (\ref{eqn:f1})-(\ref{eqn:bsj}). The coefficients $f_1$ and $f_2$ are associated with the mean-motion resonance while $\CA$ and $\CB$ describe secular interactions. The value of $f_2$ listed for the 2:1 includes the contribution of the indirect potential. }
\label{tab:coeff}
\end{table} 

Terms involving $t_e$ and $t_a$ describe eccentricity damping and orbital migration, respectively, due to disc torques. We assume that eccentricity damping alone conserves the planet's orbital angular momentum by setting the coefficient $p=3$ (see also section 2.1 of \citealt{goldreich_schlichting_2014}). To ensure convergent migration, we set $t_a < 0$ for planet 1, and do not allow planet 2 to migrate. In reality both planets probably migrate, but their mutual interaction is controlled by the relative migration rate. 

The outer resonant planet 
obeys an analogous set of equations:
\begin{align}
    \dot{n}_2 =&  -3(q+1)n_2^2\mu_1\left(e_1 |f_1|\sin\phi_1 - e_2|f_2|\sin\phi_2\right)\nonumber \\
   &+ \frac{pn_2e_2^2}{t_{e}} \label{eqn:n2dot} \\
    \dot{e}_2 =& -n_2 \mu_1 |f_2| \sin \phi_2 \nonumber \\ 
    &+ n_2\mu_1 |\CB(\alpha_{12})|e_1 \sin(\varpi_1 - \varpi_2)    \nonumber  \\
    &- n_2\alpha_{23}\mu_3|\CB(\alpha_{23})|e_3 \sin (\varpi_2 - \varpi_3) \nonumber  \\   
    &-\frac{e_2}{t_{e}} \\
    \dot{\varpi}_2 =& \,\,\frac{n_2\mu_1|f_2|}{e_2}\cos\phi_2  \nonumber \\ 
     &+ n_2\mu_1 \left[ 2|\CA(\alpha_{12})| - |\CB(\alpha_{12})|\frac{e_1}{e_2}\cos(\varpi_1 - \varpi_2) \right] \nonumber  \\ 
     &+ n_2\alpha_{23}\mu_3\left[ 2|\CA(\alpha_{23})| - |\CB(\alpha_{23})|\frac{e_3}{e_2}\cos(\varpi_2 - \varpi_3) \right]  \\ 
    \dot{\phi}_2 =& \,\,(q+1)n_2 - qn_1 - \dot{\varpi}_2  \,.\label{eqn:phi2dot} 
\end{align}
As mentioned above, we have removed the migration term involving $t_a$ for planet 2 since outcomes depend only on the relative migration rate. For simplicity, the eccentricity damping time $t_e$ is assumed equal for planets 1 and 2.

\rev{We set $t_a = -10^7\,n_1^{-1}$ and remove all terms due to disc torques once the pair has equilibrated in resonance. For our simplifying assumptions, the final equilibrium state depends on the ratio 
$t_e/|t_a|$ \citep{goldreich_schlichting_2014, terquem_papaloizou_2019, choksi_chiang_2020}.
A common estimate is $t_e/|t_a| \sim (h/a)^2 \sim 10^{-3}$ for disc aspect ratios $h/a \sim 0.03$ at $a \sim$ 0.3 AU \citep{tanaka_etal_2002, tanaka_ward_2004, cresswell_nelson_2008}. 
This estimate for $t_e/|t_a|$ derives from idealized discs with power-law surface density profiles and constant temperatures.
Accounting for more realistic disc properties, including gaps, horseshoe substructures, and circumplanetary material (e.g., \citealt{dangelo_etal_2003, dangelo_etal_2005}; \citealt{masset_etal_2006}; \citealt{duffell_chiang_2015}), and non-isothermal equations of state (\citealt[][their fig.~3]{kley_nelson_2012}), can change both $t_e$ and $t_a$.}

\rev{Given these complications, we do not rely on the disc's aspect ratio $h/a$ to set $t_e/|t_a|$. Instead, we use equations (49)--(51) of \cite{terquem_papaloizou_2019} to choose, for each integration, the value of $t_e/|t_a|$ that yields an equilibrium $\Delta$ (fractional distance from period commensurability) of 1\%, close to the values observed for {\it Kepler} sub-Neptunes. The precise value depends on the individual planet masses and the resonance being considered. For reference, when $\mu_1 = \mu_2 = 3 \times 10^{-5}$, we calculate that $t_e/|t_a| = 1 \times 10^{-4}$ for the 3:2 resonance and $4 \times 10^{-5}$ for the 2:1. These are  similar to the values advocated by \citet[][their fig. 7]{huang_ormel_2023}. Our results for TTV phases do not seem especially sensitive to $t_e/|t_a|$. We find for the example model in Fig.~\ref{fig:rep} that changing $t_e/|t_a|$ from its nominal value by a factor of 10 changes the TTV phase proxy $\mdd$ by a factor of 2.}

The resonantly forced eccentricities used in
equations
(\ref{eqn:dt_full})-(\ref{eqn:dt_full2}) 
to determine TTVs analytically 
are
\begin{align}
    e_{\rm forced,1} &= \frac{|f_1|\mu_2}{(q+1)\Delta \sqrt{\alpha_{12}} } \label{eqn:eforced1} \\ 
 e_{\rm forced,2} &= \frac{|f_2|\mu_1}{(q+1)\Delta} \,,
\end{align}
obtained by taking $\dot{\phi}_1 = \dot{\phi}_2 = 0$ while neglecting all but the resonant interaction terms.

Finally, we account for the secular back-reaction of planets 1 and 2 onto planet 3:
\begin{align}
\dot{n}_3 =& \,\,0\\
    \dot{e}_3 =& \,
    \,n_3\mu_1 |\CB(\alpha_{13})|e_1 \sin(\varpi_1 - \varpi_3)  \nonumber \\ 
    &+ n_3 \mu_2 |\CB(\alpha_{23})|e_2 \sin (\varpi_2 - \varpi_3) \\ 
    \dot{\varpi}_3 =& \,\, n_3\mu_1 \left[2|\CA(\alpha_{13})| - |\CB(\alpha_{13})|\frac{e_1}{e_3}\cos(\varpi_1 - \varpi_3) \right] \nonumber \\ 
    &+n_3\mu_2 \left[ 2|\CA(\alpha_{23})| - |\CB(\alpha_{23})|\frac{e_2}{e_3}\cos(\varpi_2 - \varpi_3)\right] \,. \label{eqn:w3dot}
\end{align}
Note that we neglect eccentricity damping and migration of planet 3. To generate non-zero TTV (transit timing variation) phases for planets 1 and 2, we need planet 3's orbit to be sufficiently eccentric, but do not model explicitly how planet 3 acquired or retained its eccentricity.

We numerically integrate the 10 coupled differential equations in (\ref{eqn:n1dot})-(\ref{eqn:w3dot}) using the \texttt{lsoda} algorithm implemented in the \texttt{solve-ivp} module of \texttt{scipy} \citep{virtanen_etal_2020}. The fractional tolerance of the solutions is set to $10^{-9}$.

\subsection{Checks}
\label{subsec:assumptions}
Equations (\ref{eqn:dt_full})--(\ref{eqn:Z}) for the TTVs of a near-resonant pair of planets are derived from \cite{lithwick_etal_2012} who assumed that the planets' mean longitudes do not deviate from linear ephemerides by more than $\sim$1 rad (see their equations A7 and A22, and their appendix A.2). This requires $\Delta$, the pair's distance from a period commensurability, to be not too small. In our paper we have focussed on resonant pairs of planets forced by third bodies to have free eccentricities $\gg$ forced eccentricities, and to be apsidally aligned with $\varpi_1 \simeq \varpi_2$; under these conditions, the assumption of \cite{lithwick_etal_2012} translates to $\Delta \gtrsim \sqrt{\mu |e_1 - |f_2/f_1|e_2|}$, where $\mu$ is either $\mu_1$ or $\mu_2$. Our modeled sub-Neptunes have $\Delta \sim 1\%$, $e \lesssim 0.1$, and $\mu \sim 3 \times 10^{-5}$, and thus satisfy this requirement.

As an added check, we integrate the same representative system shown in Figure \ref{fig:rep} using the \texttt{REBOUND} $N$-body code \citep{rein_liu_2012}. We use the WHFast \citep{rein_tamayo_2015} implementation of the \cite{wisdom_holman_1991} algorithm with a timestep of 0.02$P_1$, where $P_1$ is the orbital period of the innermost planet. Migration and eccentricity damping forces are included using the \texttt{modify-orbits-forces} package in \texttt{REBOUNDx} \citep{tamayo_etal_2020}. Simulation outputs are recorded every $500 P_1$ using a \texttt{SimulationArchive} \citep{rein_tamayo_2017}. 
We compute TTV phases by re-running segments 
of our simulation with finer timestepping. For each snapshot in the \texttt{SimulationArchive} we integrate forward by $300P_1$ and record transit times to a precision of $10^{-7}P_1$ using the bisection algorithm suggested in the \texttt{REBOUND} documentation.\footnote{\url{https://rebound.readthedocs.io/en/latest/ipython_examples/TransitTimingVariations/}} Transiting conjunctions (TCs) are identified using the same method applied to the observations in Figure \ref{fig:ttv_data2}, and used to compute TTV phases. If no TCs occur in the given 
segment, we move on to the next simulation snapshot. 
We have checked that we obtain nearly identical results for the TTV phases if instead of using TCs as our reference we use the times when the two planets have true longitudes that differ from zero by less than 0.1 rad. 

\begin{figure*}
\includegraphics[width=0.7\textwidth]{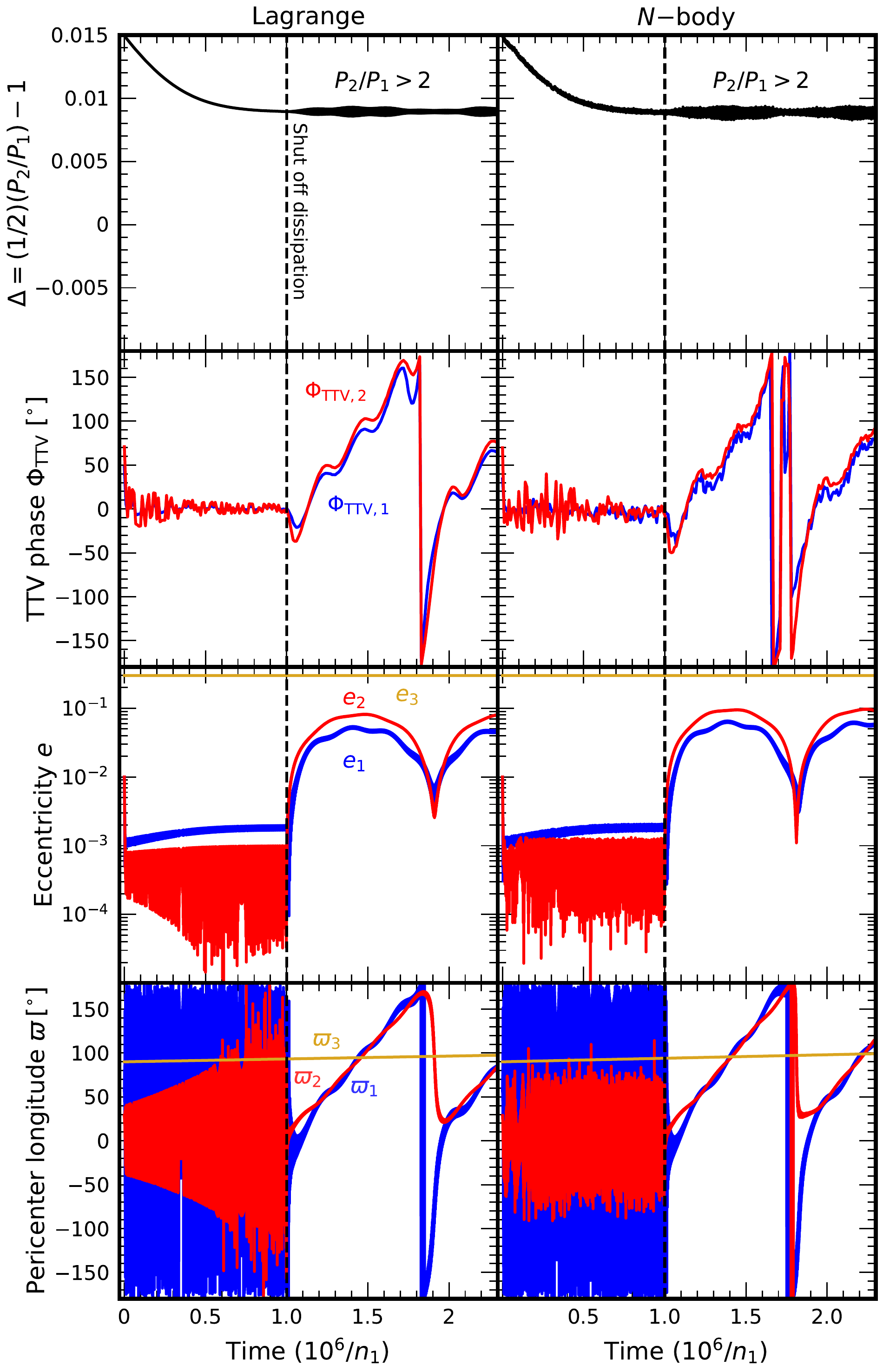} \\ 
\caption{Comparison of results obtained using Lagrange's equations of motion (\ref{eqn:n1dot})-(\ref{eqn:w3dot})  supplemented by analytic TTV formulae (\ref{eqn:dt_full})-(\ref{eqn:Z}) (left column) against results from a direct $N$-body simulation (right column). Initial conditions are the same as in Figure \ref{fig:rep}. 
}
  \label{fig:Nbody1}
\end{figure*}

Figure \ref{fig:Nbody1} compares solutions obtained from our standard method of integrating Lagrange's equations with those from \texttt{REBOUND}.

\section{Differences between the 3:2 and 2:1 resonances}
\label{sec:appendix_B}
In section \ref{sec:params} we showed that TTV phases are easier to excite for planets near the 2:1 than near the 3:2 (Fig.~\ref{fig:parameter_survey}). How the degree of phase excitation depends on mass ratio $m_1/m_2$ also differs between the two resonances (Fig.~\ref{fig:mratio}). Here we explore the reasons for these behaviours.

As discussed in section \ref{subsec:dissect}, a non-zero TTV phase shift depends on a time-varying mean motion. After disc torques are removed, and the third body enforces $\varpi_1 = \varpi_2$ so that $\phi_1 = \phi_2$, the only way to generate mean motion variations in planets 1 and 2 is for their eccentricities to deviate from the ratio 
$(e_2/e_1)_{\rm res} = |f_1/f_2|$ (see equations \ref{eqn:n1dot} and \ref{eqn:n2dot}). Deviations from $(e_2/e_1)_{\rm res}$ are easier to achieve at larger $\Delta$, i.e. for systems that behave more secularly than resonantly. In the secular limit, planets 1 and 2 are driven by disc eccentricity damping into a secular eigenmode --- a fixed point in the secular theory, where the eccentricities do not change and the apses precess at the same rate (for given $\alpha_{12}$), distinct from the resonant fixed point discussed elsewhere in this paper. The relevant mode has $\varpi_1 = \varpi_2$ and is easiest to write down in the test particle limit: when $m_1 \ll m_2$, $(e_2/e_1)_{\rm sec} = |2\CA/\CB|$, and when $m_2 \ll m_1$, 
$(e_2/e_1)_{\rm sec} = |\CB/2\CA|$.

\begin{figure*}
\includegraphics[width=0.99\textwidth]{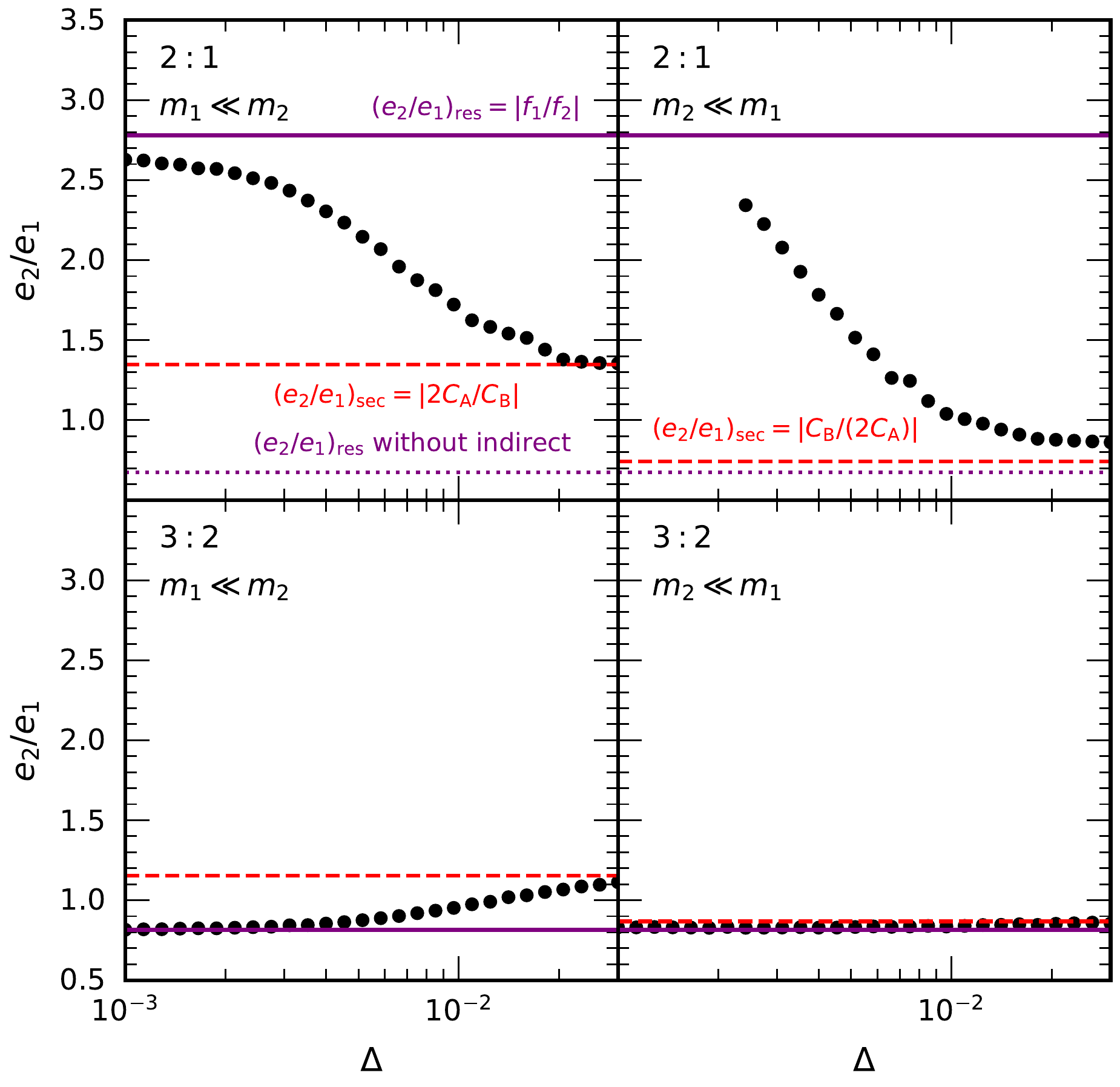} \\ 
\caption{ 
\rev{TTV phases are easier to excite for the 2:1 resonance than for the 3:2.} The closer $e_2/e_1$ is to $(e_2/e_1)_{\rm res} = |f_1/f_2|$, the harder it is to generate non-zero $\dot{n}_1$ and $\dot{n}_2$ from the resonant interaction (assuming $\phi_1 = \phi_2$ in equations \ref{eqn:n1dot} and \ref{eqn:n2dot}, a condition enforced by planet 3 in our simulations post-disc damping), and thus the harder it is to generate non-zero TTV phases.
For smaller $\Delta$, the system  (modeled for this figure in the  $m_1 \ll m_2$ and $m_2 \ll m_1$ test particle limits) is driven by disc eccentricity damping to a resonant fixed point where $(e_2/e_1) = (e_2/e_1)_{\rm res}$ (purple solid line). For larger $\Delta$, the resonance weakens and the system behaves more secularly; it is driven to a secular fixed point where $(e_2/e_1) = (e_2/e_1)_{\rm sec}$ (red dashed line), which takes different forms depending on which test particle limit obtains. The resonant and secular fixed-point eccentricity ratios differ more for the 2:1 resonance than the 3:2, partly because the 2:1 resonance has an indirect potential that decreases $f_2$ (dotted purple line). Thus there is more parameter space for mean motion variations and by extension non-zero TTV phases for the 2:1 than the 3:2. In the top right panel we omit simulation data at $\Delta < 0.002$ that suffered from numerical instabilities.}
  \label{fig:Delta_eratio}
\end{figure*}

Figure \ref{fig:Delta_eratio} confirms that post-disc damping, $e_2/e_1$ varies between the resonance-dominated limit $(e_2/e_1)_{\rm res}$ at small $\Delta$, and the secular-dominated limit $(e_2/e_1)_{\rm sec}$ at larger $\Delta$. The figure derives from simulations like the one in Figure \ref{fig:rep}, but performed in the $m_1 \ll m_2$ and $m_2 \ll m_1$ test particle limits, and with varying $\te/|\ta|$ to generate different equilibrium $\Delta$ values. The values of $e_2/e_1$ plotted are measured post-disc, at the peak of the secular cycle driven by planet 3; we found that this ratio hardly changed throughout the cycle.

We see from Figure \ref{fig:Delta_eratio} that 3:2 pairs are confined to a much narrower range of $e_2/e_1$ than 2:1 pairs. The restriction is particularly severe for 3:2 resonance in the $m_2 \ll m_1$ limit (bottom right panel).
The smaller the range of permitted values, the closer $e_2/e_1$ is to $(e_2/e_1)_{\rm res}$, and the harder it becomes to excite TTV phases. Thus we can make some sense of the trends seen in Figures \ref{fig:parameter_survey} and \ref{fig:mratio}. The 2:1 resonance is more susceptible to secular forcing than the 3:2, in part because the 2:1 is weakened ($|f_2|$ is made smaller) by the indirect potential.

\bsp	
\label{lastpage}
\end{document}